# DC Cancellation As a Method of Generating a $t^2$ Response and of Solving the Radial Nonobservability Problem in a Concentric Free-Falling Two-Sphere Equivalence-Principle Experiment in a Drag-Free Satellite

Benjamin Lange
1922 Page St.
San Francisco, CA 94117-1804
Tel. and Fax 1-415-221-6600, Ext. 310
email: blange@virtualpbx.com

March 1, 2001

## Abstract

This paper presents a new method for doing a free-fall Equivalence-Principle (EP) experiment in a satellite which solves two 25-year-old problems that have previously blocked this approach. By using large masses to change the gravity-gradient at the proof masses, the orbit dynamics of a drag-free satellite may be changed in such a way that the experiment can mimic a free-fall experiment in a constant gravitational field on the earth. An experiment using a sphere surrounded by a spherical shell both completely unsupported and free falling has previously been impractical because: 1) it is not possible to distinguish between a small EP violation and a slight difference in the semi-major axes of the orbits of the two proof masses and 2) the position difference in orbit due to an EP violation only grows as $t$ whereas the largest disturbance grows as $t^{3/2}$. Furthermore, it has not been known how to independently measure the positions of a shell and a solid sphere with sufficient accuracy. The measurement problem can be solved by using a two-color transcollimator (see main text); and since the non-observability and $t$-response problems arise from the earth's gravity gradient and not from its gravity field, one solution is to modify the earth's gravity gradient with local masses fixed in the satellite. Since the gravity-gradient at the surface of a sphere, for example, depends only on its density; the gravity gradients of laboratory masses and of the earth unlike their fields are of the same order of magnitude. In a drag-free satellite spinning perpendicular to the orbit plane, two fixed spherical masses whose connecting line parallels the satellite spin axis can generate a DC gravity gradient at test masses located between them which cancels the combined gravity gradient of the earth and differential centrifugal force. With perfect cancellation, the non-observability problem vanishes and the response grows as $t^2$ along a line which always points toward the earth, i.e. in the radial direction. In the practical case where the cancellation is not perfect, the $t^2$-response can hold for about $10^4$ to $10^6$ seconds (depending on the cancellation accuracy), and the non-observability is also suppressed proportional to the accuracy of the cancellation. Experience with a prior drag-free satellite indicates that this cancellation may be accomplished with an accuracy of at least $10^{-2}$ and possibly $10^{-4}$ to $10^{-6}$, and this ameliorates the measurement problem sufficiently that equivalence-principle tests with accuracies between $10^{-20}$ and $10^{-23}$ $g$ may be possible. $t^2$-response times between $10^5$ and $10^6$ seconds are equivalent to a very tall (0.1 to 10 AU) drop tower with an effective value of $g$ equal about 3.4 meters/sec² which is 3/7 of the value of $g$ at the orbital altitude.

## Introduction

There is a major problem with any satellite-based Equivalence-Principle (EP) experiment, and there is a second significant problem with any free-fall version. The problem common to all satellite experiments is that a violation of the Equivalence Principle cannot be distinguished from a very small difference in the semimajor axes of the two test masses. It will be shown later that this position tolerance is so small that high-precision EP measurements in a satellite are essentially impossible without some kind of countermeasure. The second problem for free-fall tests arises from the fact that an EP violation causes the two proof masses to separate proportional to $t$ not $t^2$. It turns out that the largest disturbance in the experiment is random walk due to collisions from the gas molecules in the residual vacuum. This disturbance grows as $t^{3/2}$ so that the experiment error would worsen as $t^{1/2}$ rather than improving with increased averaging time. This paper will present a method of modifying the orbit dynamics in a drag-free satellite version of a free-fall EP experiment which solves both of the above problems.

Because of recent developments in string and related gravitation theories, a high-precision EP test is one of the most important experiments yet to be done. Damour [1] has summarized this exceptionally well. As a further example, two recent papers [2, 3] have calculated the expected size of an EP violation caused by string dilatons. Reference [2] has shown that under the assumption of universality of the dilaton coupling functions, string-loop corrections allow for the existence of common local minima to which the dilaton field would be attracted as the universe expands. If the deviation from this minimum were of order unity at the beginning of the radiation era, this assumption leads to the conclusion that a violation of the equivalence principle should lie in the range of $10^{-12}$ to $10^{-24}$ $g$. An initial deviation of order unity, however, does not include the effects of inflation; and given that quantum creation of dilatons restored at least part of the field diluted by inflation, References [2] and [3] together show that the expected violation should be between $10^{-26}$ and $10^{-38}$ $g$. In addition to estimating the expected range of values of an EP violation, Reference [2] calculates a relation between the Eddington PPN parameter, $\gamma$, and a violation of the equivalence principle, $\Delta a / a$. If the result, $\Delta a / a \approx 10^{-5}(1 - \gamma)$, is correct; then a test of the equivalence principle is a much more sensitive method of detecting a scalar field than experiments which measure $\gamma$ or $1 - \gamma$ directly such as the Relativity Gyroscope or the gravitational bending of electromagnetic waves. Motivated by these results and by what is possible, this paper will investigate the possibility of designing an EP experiment in an accuracy range of $10^{-20}$ to $10^{-23}$ $g$.

One obvious design meeting the above requirements would be a drag-free satellite [4] with two free-falling concentric spherical proof masses, a solid sphere inside of a spherical shell. Not only is this configuration of proof masses the natural way to design an experiment, but it is simple and relatively easy to realize. Until now, however, this design has not been practical because of the problems discussed above. In addition it has initial-condition problems which have been discussed in Jafry [5]. Because of these difficulties, prior satellite EP designs have used single-axis force-rebalance differential accelerometers[1].

---

[1] This is the configuration of the STEP experiment [6, 7]. In this arrangement an EP violation causes a rebalance signal which is periodic once per satellite revolution with respect to the earth whereas miscentering causes a signal that is periodic twice per revolution. Thus it is possible for a servo to center the two cylindrical proof masses

A third problem, measuring the positions of the sphere and the shell, can be solved with a two-color transcollimator [8]. A transcollimator is an autocollimator with only one minor change, the focal length of the collimating lens is altered to focus the output beam to the center of a reflecting sphere. When this is done, a translation, $x$, of the sphere perpendicular to the beam is equivalent to a rotation, $\theta$, of the mirror of an equivalent autocollimator given by $\theta = x/a$ where $a$ is the radius of the sphere. The two-color version of the transcollimator can measure the position of the two spheres with a noise equivalent translation of $10^{-12}$ meters/Hz$^{1/2}$ $\times$ $(1\ \mu\text{watt}/W)^{1/2}$ where $W$ is the power in each light beam [8]. For example, a beam power of 100 microwatts would give $10^{-13}$ meters/Hz$^{1/2}$. It will turn out, however, that the accuracy of the experiment is not determined by the readout noise; and it will be convenient in some cases to reduce the beam power considerably below one microwatt. This paper will assume that the measurement problem has been solved, and will concentrate on the problems of the $t$-dependence and radial non-observability.

Figure 1 shows the experimental setup. The satellite will use the inner solid sphere as the drag-free control reference. Since only one mass can be drag free, the outer shell must be controlled by applying a force to its surface until the experiment begins when it will be set free. Prior to the experiment, the spherical shell would be forced by electrodes similar to an electrically supported gyroscope. The basic experiment involves aligning the two spheres as accurately as possible and then releasing the outer shell to let the two concentric spheres free fall for the duration of the experiment.

Both spheres and the satellite would spin with their spin axes normal to the orbit plane, but the shell should spin in the opposite direction as the sphere and the satellite which should not have exactly the same spin rate. Spinning both the proof masses and the satellite is necessary to attenuate some of the disturbances, and it provides a defined orientation of the proof masses perpendicular to the experiment plane. The spin of each body will be very rapid, of the order of one Hz. Rapid spin is possible in this setup because the test masses are free floating eliminating centrifugal-force errors due to miscentering; and rapid spin is important because, in addition to greatly reducing temperature gradients, any errors which occur during the first half cycle of roll are never removed so that many cycles are necessary to make any such error small in comparison to what it would have been in the absence of roll.

This paper is a more detailed discussion of a very short paper presented at the Eighth Marcel Grossmann Conference in Jerusalem [9].

---

using the twice per revolution error signal. In addition any residual centering error signal can be separated in frequency from the EP signal. Because of the force-rebalance servo, the proof masses are not separately free-falling, and the problem that the separation only grows as $t$ does not arise. In contrast, a free-fall solution has no forces or torques deliberately applied to the proof masses and the radial-nonobservability and $t$-response problems are solved by the method explained in this paper.

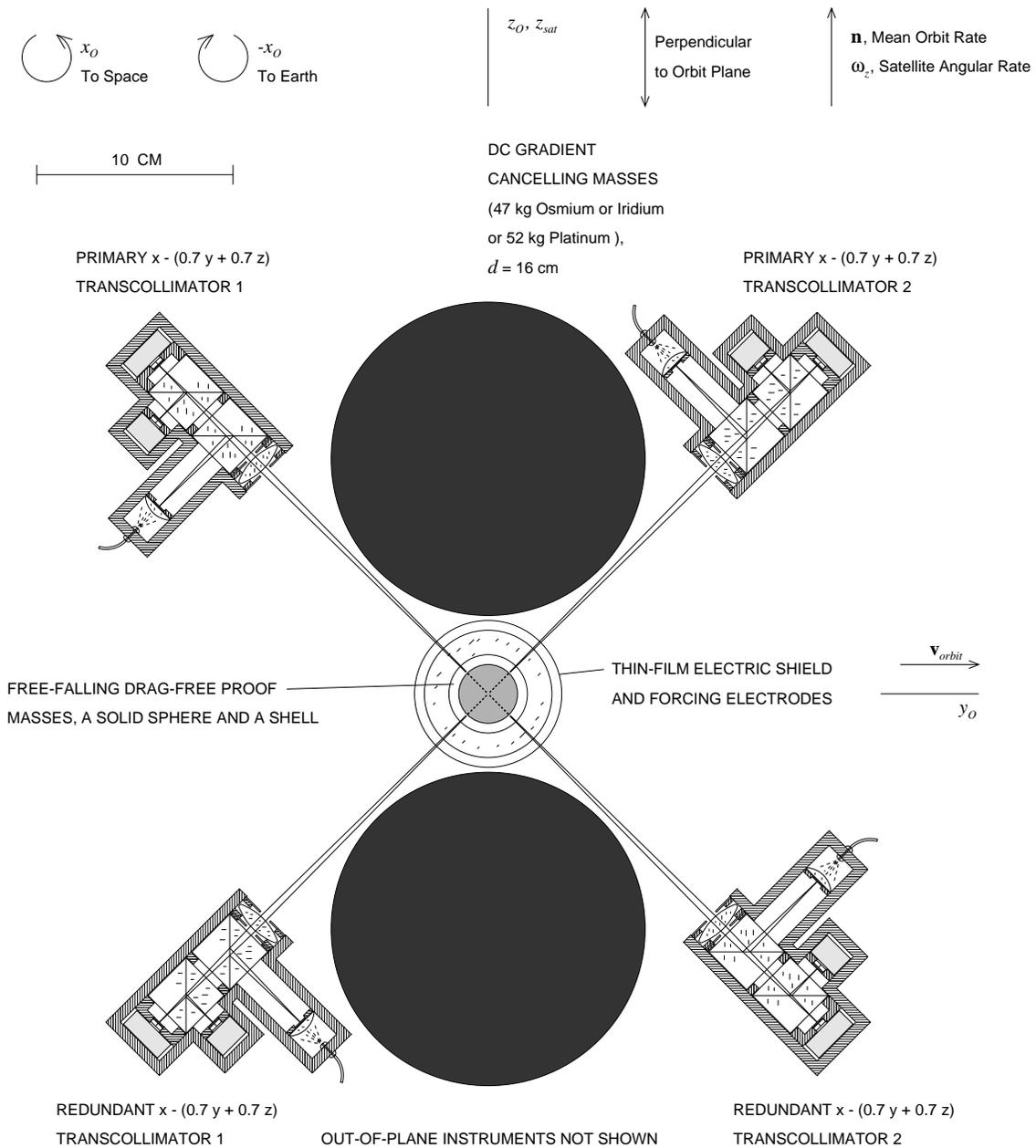

Figure 1. DC Cancellation of the Radial-Offset Error

## Section 1.  The Radial Non-Observability Problem

That there is a fundamental problem of observability in an equivalence-principle experiment involving two point masses in orbit may be seen directly by considering two masses with different gravitational constants $\mu$ and $\mu'$ (See Appendix A).  Assume that the two masses are in circular orbits at the same radius.  Because $\mu'$ for the primed mass is slightly different from $\mu$ for the other mass, the difference $\Delta\mu \equiv \mu' - \mu$ will cause the two mean angular rates, $n'$ and $n$, to be different.  Thus $\Delta\mu$ can be measured by observing the in-track separation of the two bodies over time.  It is possible get this same rate difference in $\Delta n$, however, without any violation of the equivalence principle by changing the orbital radius of one of the masses by $\delta a = -a\,\Delta\mu\,/\,3\mu$. For $\Delta\mu\,/\,\mu$ in the range of interest for a precision equivalence-principle measurement, the change in the radius is too small to be measured. A violation of the EP and a radial offset error cannot be distinguished even in principle no matter how small the noise of the measuring instrument unless there is an independent method of accurately measuring the "gravitational centers" of the test masses.

The non-observability problem may be seen more formally from the equations of relative motion of the two masses in a locally-level reference frame. If the satellite gravity gradients in Equation B1 are neglected, the linearized equations of relative motion for slightly different values of $\mu$ are adequately approximated by the usual Euler-Hill equations ($\ddot{x} - 3n^2 x - 2n\dot{y} = f_x$ and $2n\dot{x} + \ddot{y} = 0$) driven by a small constant specific force, $f_x$, equal to the violation of the equivalence principle. The experiment plane is $x$-$y$; and $x$ and $y$ are the relative offsets between the two spheres in the radial and orbit-velocity directions respectively. By solving the equations for an initial $x_0$ and a nonzero $f_x$,

$$x(t) = x_0 + (f_x + 3n^2 x_0)(1 - cnt)/n^2 \qquad (1)$$

$$y(t) = -2(f_x + 3n^2 x_0)(nt - \mathcal{S}nt)/n^2. \qquad (2)$$

where $\mathcal{S}$ and $c$ are sine and cosine. A nonzero $f_x$ could be detected after sufficient time by the drift in $y$. The problem is that the combination $f_x + 3n^2 x_0$ always occurs together in Equation 2 so that a drift in the y-axis could either come from a nonzero $f_x$ or from an unknown value of $x_0$. Interesting values of $f_x$ correspond to very small values of $x_0$. As an example if $f_x = 10^{-20}\,g$, it could be mimicked by $x_0 = 3 \times 10^{-14}$ meters. Such a small value is not absolutely measurable; and in fact since the best manufacturing accuracies of spheres is of the order of $10^{-8}$ meters, the centers are not even defined to this accuracy.

Two ways of solving the problem of radial non-observability are gravity-gradient cancellation and gravity-gradient excitation of the difference of the "gravitational centers" of the two spheres. The latter method does not work very well and will not be discussed here, but is the subject of a separate paper [10]. The idea of modifying the gradient of the gravity field in a drag-free satellite by specially placed masses was first applied in the 1972 Stanford-University drag-free flight, Triad I or DISCOS [11], where the system was designed to meet both a $g$ and a gravity-gradient specification. To achieve the specifications a compensating mass was added, and in addition the fuel tanks were built in the form of two toroidal rings symmetrically placed with respect to the cavity. As originally pointed out by Richard VanPatten, when the ratio of diameter to separation is correctly chosen, their gravitational field and gravity gradient are canceled independently of the mass of the gas in the tanks; and these are known as "gravitational Helmholtz coils". In 1982 Robert Forward published a technique which flattens space-time using a locally-level ring of six masses [12]. In principle Forward's scheme could be used in a satellite controlled to the local vertical to cancel the earth's gravity gradient, but this is not practical for a high precision equivalence-principle experiment since rapid satellite spin is necessary to reduce many of the disturbances and because $g$ rotates once per orbit so that the response to an EP violation only grows as $t$ and not $t^2$. The AC version of the technique described in this paper, however, surrounds the cavity with a toroidal mass whose symmetry axis is in the orbit plane and which spins about an axis in the plane of the torus perpendicular to the orbit plane to combine the two methods referred to above into one system; but that is also covered in [10].

## Section 2. Solution by DC Gravity-Gradient Cancellation

Figure 1 shows the design of a two-sphere equivalence-principle experiment which suppresses the observability problem without having to measure $x$ absolutely to the order of $10^{-14}$ meters and which results in a $t^2$ response to an EP violation. The two large dark masses in the figure are fixed in the satellite and create a gravity gradient whose largest value is parallel to the satellite spin axis. The two proof masses, a solid sphere and a transparent shell surrounded by the electric shield and forcing electrodes, are shown in the center of the drawing at the intersection of four focused light beams from sets of two-color transcollimators which measure the positions of each sphere. Either a primary or a redundant set alone gives a full independent measurement of each sphere in three axes, and either set can provide the inputs into the satellite's drag-free control system so that the proof masses remain free floating. The inner sphere is a conducting heavy metal such as Uranium and the outer shell is a transparent conductor (see www.iivi.com for the availability and properties of transparent conductors).

The satellite spin axis is in the plane of the paper along the vertical axis, and the plane of the orbit is perpendicular to the paper with the satellite velocity from left to right. The radial direction to space is out of the paper toward the observer who is looking down toward the earth. The subscript on the coordinate labels stands for the orbit or locally-level frame, and the drawing captures the rotating satellite at the instant when the satellite axes and the locally-level axes are aligned.

It is shown in Appendix B1 that when the gravity-gradient tensor has been adjusted using compensating masses to be diagonal in satellite coordinates to the level of accuracy discussed in Section 3.2 with eigenvalues, $g_{11}$, $g_{22}$, and $g_{33}$; the relative equations of motion of the two spheres in locally-level orbit coordinates in a rotating satellite with its spin perpendicular to the orbit plane become

$$\ddot{x} - \left(3n^2 + \frac{g_{11} + g_{22}}{2} + \frac{g_{11} - g_{22}}{2}c2\right)x - \left(\frac{g_{11} - g_{22}}{2}s2\right)y - 2n\dot{y} = f_x + \Delta f_{nsx} + f_{dx} + f_{ggx} \qquad (3)$$

$$-\left(\frac{g_{11} - g_{22}}{2}s2\right)x + 2n\dot{x} + \ddot{y} - \left(\frac{g_{11} + g_{22}}{2} - \frac{g_{11} - g_{22}}{2}c2\right)y = f_{dy} + \Delta f_{nsy} + f_{ggy} \qquad (4)$$

and

$$\ddot{z} + (n^2 - g_{33})z = f_{dz} + \Delta f_{nsz} + f_{ggz} \qquad (5)$$

where $s2$ and $c2$ are the sine and cosine of two times the satellite roll angle; and the subscripts $d$, $ns$, and $gg$ are the disturbances, non-spherical $g$ fields, and the neglected gravity-gradient terms respectively. This system is linear with periodic coefficients, and the period is twice roll rate for the same reason that tides occur twice per day on the earth.

Revision 3.0

To suppress the problem of non-observability, the two dark gradient-cancellation masses in Figure 1 are chosen so that their combined gravity-gradient tensor at the center of the cavity in satellite coordinates is

$$g_{ij} = \begin{bmatrix} -3n^2 & 0 & 0 \\ 0 & -3n^2 & 0 \\ 0 & 0 & 6n^2 \end{bmatrix}. \tag{6}$$

This is possible with heavy metals such as Osmium, Iridium, or Platinum (See Appendix B1 and Equation B2). In the ideal case where the gradients from all other satellite masses are compensated so that Equation 6 holds exactly, $(g_{11} + g_{22})/2 = -3\,n^2$ and $g_{11} - g_{22} = 0$ and the $3\,n^2$ term in Equation 3 is canceled. Since $g_{11} - g_{22} = 0$, the periodic terms do not show up in the equations, and this case is referred to as DC cancellation. The advantage of DC cancellation is that the cancellation masses are smaller and also the equations can be solved exactly so that the system is easier to understand. In the ideal case with zero disturbance and perfect cancellation the x-y equations become

$$\ddot{x} - 2n\dot{y} = f_x \tag{7}$$

$$2n\dot{x} + \ddot{y} + 3n^2 y = 0. \tag{8}$$

From Appendix B2 the solutions of these equations for an initial $x_0$ and a constant specific force, $f_x$, are,

$$x = x_0 + \frac{3}{14} f_x t^2 + \frac{4 f_x}{49 n^2} (1 - c\sqrt{7}nt) \tag{9}$$

$$y = -\frac{2 f_x}{7n} t + \frac{2 f_x}{7\sqrt{7 n^2}} \, S\sqrt{7}nt \,. \tag{10}$$

where $S$ and $c$ are sine and cosine.

It can be seen that the principal term proportional to $f_x$ grows as $t^2$ in the x-equation and as $t/n$ in the y-equation, and in this case the dependence on $x_0$ and $f_x$ are not connected.[2] In the non-ideal case where the cancellation is not perfect, it is shown in Appendix C that Equations 9 and 10 still hold approximately; but the unobservability appears in the combination $f_x + k^2 x_0$ instead of $f_x$ alone where $k^2 = 3n^2 + (g_{11} + g_{22})/2$, i.e. $k^2$ is the error in the cancellation. Compared with Equations 1 and 2 where the combination is $f_x + 3n^2 x_0$ the unobservability is suppressed by $k^2/3n^2$, but in addition the $t^2$ dependence is still maintained (Cf. Appendix C3). This solution allows very large signals and arbitrarily fast roll rates with good roll averaging, but Equation (5) or B3 shows that it is unstable in $z$ with a time constant of about 450 seconds so that the $z$-axis must be controlled using the light pressure from the transcollimators.

---

[2] This corresponds to a drop tower with $g = 3/7\ g_{orbit} \approx 3.4$ m/sec². For a space equivalence-principle test to truly correspond to a free-fall Galileo-type experiment, three features are necessary: it should fall toward the center of the earth, the EP violation should grow as $t^2$, and the effective gravity should be a significant fraction of $g_0$. These conditions are not satisfied by constrained cylinders using force rebalance [6, 7] or by Eötvös ($g \approx 0.013$ m/sec²) or Dicke ($g \approx 0.006$ m/sec²) torsion balances.

## Section 3. Experiment Accuracy

The accuracy by which the equivalence principle can be tested using the method described in this paper depends on the accuracy with which the surfaces and centers of mass of the two spherical test masses can be measured, the accuracy of the gravity-gradient cancellation, the disturbing forces, and the transcollimator noise. In addition to the accuracy of the experiment, separate control of the unstable $z$-axis root will be discussed in this section because some issues of $z$-axis control are related to system accuracy.

### Section 3.1 The Accuracy with which the Centers of Mass and the Surfaces of the Spheres Can Be Measured

If it is assumed that the centers of mass of the spheres and the effective "gravitational centers" are very close, then accurately measuring the surface errors of the spheres and the locations of the centers of mass can also be used to suppress radial non-observability and to reduce the disturbing forces which act differentially on the two spheres. While it is difficult on the ground, the possibility exists for high precision measurements in orbit. During the experiment the spheres will spin with their spin vectors actively damped [13] to each sphere's maximum axis of inertia, but during the pre-measurement phase a sphere's entire surface can be scanned by the transcollimators by commanding a large polhode angle with the active damper. The spheres can typically be polished to an accuracy of about $10^{-8}$ meters, and it can be shown that surface errors of $10^{-8}$ meters will also cause deterministic errors in the translation readout by the transcollimators of the about same size. Since the noise level of the transcollimator is about $10^{-12}$ meters/Hz$^{1/2}$, these surface errors could in principle be measured to about one part in $10^4$; but this paper will only assume that it can be done to about one percent, i.e. one part in $10^2$. If the results of these measurements are fitted to a spherical-harmonic model, the surface of the sphere can be accurately measured. Since the spheres will spin about their centers of mass; and since in the drag-free mode the disturbances in the motion of the spheres will be considerably less than $10^{-12}$ meters, the centers of mass of the spheres can also be determined to the accuracy of the surface calibration from their wobble due to any center-of-mass offset. Center-of-mass-offset wobble (transformed to an inertial reference) is at the spin frequency of the proof mass whereas surface calibration errors are higher harmonics. (Furthermore in order to prevent center-of-mass wobble from saturating the dynamic range of the transcollimators, preliminary mass balancing on the ground must be done with the proof masses suspended in an electric bearing before final assembly.) The drag-free controller error can be separated out because it is a common mode error affecting both proof masses together. Thus the noise performance of the transcollimator predicts that the spherical surface and the center of mass can be measured before the start of an experiment to at least $10^{-10}$ meters and possibly to $10^{-12}$ meters. It should be cautioned, however, that this calibration might not be stable since $10^{-12}$ meters is about one hundredth of an atomic distance, and any surface contamination due to, for example, residual gas in the cavity might reduce the accuracy to about one atomic distance, $10^{-10}$ meters. It will be shown in the next section (Tables 1 and 2), that knowing the surface to even this accuracy combined with DC cancellation is sufficient to eliminate the radial non-observability error.

## Section 3.2 The Expected Accuracy of the Gravity-Gradient Cancellation

Appendix C shows the results when the gravity-gradient cancellation is not perfect as was assumed in Appendix B2. The are two ways to determine the gravity gradient to sufficient accuracy to get good cancellation: the gravity-gradient from the satellite masses at the center of the cavity may be calculated or the accuracy of the cancellation itself may be measured. Perfect calculation of the gravity-gradient cancellation is not possible for a number of reasons: there is a limit to the accuracy with which the masses and their positions are known; $G$ is not well known; the orbit is not known exactly; there are external gravity gradients from the sun, moon, $J_2$, and other gravitational harmonics; the satellite dimensions change with temperature; fluids such as the control gas or liquid Helium must be managed so that their gravity gradients remain sufficiently small; the orbit normal changes due to the precession of the orbit unless it is equatorial or polar; etc. These effects are summarized in Table 1 in this section.

The accuracy of the gravity-gradient cancellation can be estimated from prior satellite experience. When the DISCOS module for the 1972 drag-free flight [11] was designed, the gravity and gravity-gradient at the center of the cavity of every mass in the satellite were calculated before the flight [14]. Since DISCOS met its performance requirement almost exactly, these calculations are considered to be reliable; and they show that even without any attempt to influence the gradients, a typical gravity gradient at the proof mass would be of the order of $3 \times 10^{-8} / \sec^2$. After the DISCOS calculations were complete, a compensating mass was added to the system to bring the gravity field at the proof mass within the specification of $10^{-11} g$. This mass then caused the gravity gradient to rise a factor of 3 to about $10^{-7} / \sec^2$, but this value was acceptable for the accuracy of that satellite. The results before adding the compensating mass, however, show that typical gravity gradients inside of a drag-free satellite can be assumed to be about $3 \times 10^{-8} / \sec^2$ without any special compensation. Since $3 n^2$ is about $3 \times 10^{-6} / \sec^2$, the accuracy of cancellation would be about one part in 100 even if no special steps were taken. Since Figure C1 shows that with a cancellation error of $3 \times 10^{-8} / \sec^2$ the $t^2$ behavior holds for about $10^4$ seconds, DC Cancellation gives very respectable performance even if no attempt is made to improve on the DISCOS cancellation accuracy and thus guarantees that the method will work. Since no attempt was made in DISCOS to improve the gravity-gradient cancellation (with compensating masses, for example), it may be possible (with cancellation error measurement) to improve on DISCOS by a factor of 100 or even $10^4$. Table 1 shows the cancellation errors under the assumption of no improvement on DISCOS, an improvement of 100, and an improvement of $10^4$; and Table 2 shows the resulting performance and experiment error. The last column includes a low-temperature version of the experiment, and thus includes Helium tides in the cancellation errors. The disturbance from the charge on the inner sphere is proportional to $x$ and is also included in the cancellation errors. The charge on the shell is included in the disturbances in Table E1. For the shell and gg-mass gravity-gradient errors, the error mass is assumed to be about one percent of the total mass in a shell of the thickness of the polishing errors.

| Cancellation Improvement | | Fraction Beyond DISCOS | 1 | 0.01 | 0.0001 |
|---|---|---|---|---|---|
| Source of gg Cancellation Error, $k^2$ | Formula | Auxiliary Formula, Critical Values Comments, Etc. | $k^2$, 1/sec² | $k^2$, 1/sec² | $k^2$, 1/sec² |
| Satellite gg | DISCOS + | | 3.0E-08 | 3.0E-10 | 3.0E-12 |
| gg-Mass Separation Err | $9n^2 \delta d_{gg} / d_{gg}$ | $d_{gg} = 0.15$ m, $\delta d_{gg} = 10^{-4}, 10^{-6}, 10^{-8}$ m | 9.7E-09 | 9.7E-11 | 9.7E-13 |
| Thermal Drift | $9n^2 \alpha_{lin} \delta T$ | $\alpha_{ln.\ Stn} = 5 \times 10^{-7},\ 10^{-7}$/K | 5.8E-12 | 5.8E-12 | 1.2E-12 |
| Charge on Inner Sphere | $\frac{1}{2} V_1^2 \frac{dc_{11S}}{dx_{ce}} \frac{1}{x_{ce}}$ | $V_1$=10, 1 mV, $a$=1.5, $b$=2.0, gap=0.5 cm | 3.2E-11 | 3.2E-11 | 3.2E-13 |
| Helium Tides | $2Gm_{He} / d_{He}^3$ | $m_{He}$=0.63 gm, $d_{He}$=0.5 m | 0 | 0 | 3.4E-13 |
| Shell gg | $\dfrac{2Gm_{err}}{r_{sh}^3}$ | $m_{err} = \frac{4}{3} \pi r_{sh}^2 d_{err} \rho / 100$ $r_{sh}$=0.026, $d_{err}$=$10^{-8}$ m | 1.5E-14 | 1.5E-14 | 1.5E-14 |
| gg-Mass gg Error | $\dfrac{2Gm_{err}}{d_{gg}^3}$ | $m_{err} = \frac{4}{3} \pi r_{gg}^2 d_{err} \rho / 100$ $d_{gg}$=0.12, $d_{err}$=$10^{-8}$ m | 1.3E-14 | 1.3E-14 | 1.3E-14 |
| $J_2$ Error | $12 G m_e R_e^2 (\delta J_2) / r_s^5$ | $\delta J_2 = 10^{-5} J_2$ | 1.5E-13 | 1.5E-13 | 1.5E-13 |
| Gravity Harmonics | $J_{2err} \times 10^{-3} \times 10^{-2}$ | $J_{2err} = 12 G m_e R_e^2 J_2 / r_s^5$ | 1.5E-13 | 1.5E-13 | 1.5E-13 |
| Orbit Position Error | $n^2 \delta r_s / r_s$ | $\delta r_s = 10$ cm | 1.9E-14 | 1.9E-14 | 1.9E-14 |
| Sun gg | $2Gm_{sun} / d_{sun}^3$ | | 7.9E-14 | 7.9E-14 | 7.9E-14 |
| Moon gg | $2Gm_{moon} / d_{moon}^3$ | | 1.8E-13 | 1.8E-13 | 1.8E-13 |
| Cancellation RSS Error | | | 3.2E-08 /sec² | 3.2E-10 /sec² | 3.4E-12 /sec² |

Table 1. Summary of Cancellation Errors

Concerning the fluids, at room temperature VanPatten's "gravitational Helmholtz coils" can be used to make the gravity-gradient at the cavity independent of the mass of the control gas. For the low-temperature version of the experiment liquid Helium also does not seem to be a serious problem. Using the tidal mass in the MiniSTEP Study [6], 0.63 gm, and a distance of 0.5 m; the gravity gradient of the field from a Helium tide of this amplitude would be $3 \times 10^{-13} / \text{sec}^2$ so that the cancellation error, $k^2 / 3 n^2 = 10^{-7}$. By forming the reserve Helium tanks as gravitational Helmholtz coils, by moving them farther away from the cavity, and by rapidly spinning the satellite which provides artificial gravity; it should possible to manage the variations in the gravity gradient due to the Helium liquid.

Appendix C1 and Figure C1 show how the cancellation might be measured given movable trim masses. If $k^2$ is greater than zero, Equations C6 or C7 show that the response ultimately grows with an unstable time constant of $1/a$, and Figure C1 shows the length of time before the unstable response dominates the $t^2$ response. Since the unstable signals ultimately become very large, the unstable time constant can be measured; and Equations C4 and C5 show that $k^2 = 7\,a^2/3$. The measurement of $k^2$ would then be done by initially setting movable trim masses to give a relative large positive value of $k^2$, and then the trim masses would be backed off and the experiment rerun between $10^4$ and $10^6$ seconds (depending on the desired accuracy) until the unstable root was seen. If $10^6$ seconds were required, $k$ would be about $10^{-6}/\sec$ and $k^2$ about $10^{-12}/\sec^2$.

| Cancellation Improvement | | Fraction Beyond DISCOS | 1 | 0.01 | 0.0001 |
|---|---|---|---|---|---|
| | | | | | |
| Consequences of gg Cancellation Error, $k^2$ | Formula | Critical Values and Comments | $k^2 =$ 3.2E-08 /sec$^2$ | $k^2 =$ 3.2E-10 /sec$^2$ | $k^2 =$ 3.4E-12 /sec$^2$ |
| | | | | | |
| | | | | | |
| Radial Non-Observability | $k^2 x_0 / g_e$ | $x_0 = 10^{-10},\ 10^{-10},\ 10^{-11}$ meters | 3.2E-19 $g$'s | 3.2E-21 $g$'s | 3.4E-24 $g$'s |
| Centering Error Box | | Centering tolerance between gg masses | 6 mm | 0.6 mm | 0.06 mm |
| | | | | | |
| Time in $t^2$ Domain | Figure C1 | | 1.0E+04 sec | 1.0E+05 sec | 1.0E+06 sec |
| Equivalent Drop Distance | $\frac{3}{14} g_e t^2$ | 1 AU $= 1.4 \times 10^8$ km | 2.1E+05 km | 2.1E+07 km | 2.1E+09 km |
| Overall Exp. Error at $t$ | Figures 2, 3 | | 3E-19 $g$'s | 1E-19 $g$'s | 4E-23 $g$'s |

Table 2. Consequences of the Cancellation Accuracy

Since a mass-trim system is necessary in any event to keep the center of mass of the satellite at the center the cavity and to keep the principal axes of inertia aligned, it will be assumed that a second trim-mass system can be used to actively balance the cancellation by adjusting the distance between the two gravity-gradient masses. Thus it can be expected that $k^2/3\,n^2$ can be held to at least $10^{-4}$, and there is a chance that $10^{-6}$ may be possible. The gg-mass separation error in Table 1 shows how accurately changes in the distance between the two gravity-gradient masses must be realized to use the results of a cancellation-error measurement to achieve the corresponding accuracy. There is no conflict between this number and the centering error-box lower in Table 2. This gives the error requirement for the location of the sphere and shell between the gravity-gradient masses. It is much larger than the error in the separation between the masses because the derivative of the gravity gradient due to the gg masses cancels at the central point between them so that the box size is proportional to the square of the centering error whereas the gg-mass separation error is proportional to the first power.

## Section 3.3 Disturbing Forces which Act Differentially on the Proof Masses

A preliminary analysis of the disturbing forces which act on the spheres has been done. They include gravitational interactions from proof and gravity-gradient mass imperfections, magnetic forces, electric charge, gas brownian motion, large particle collisions, the radiometer effect, differential radiation pressure, differential cavity pressure from out-gassing and vacuum pumping, trapped radiation, iron cosmic rays, solar flares, proton cosmic rays, transcollimator light pressure, UV charge-control light pressure, and (for low temperatures) Helium tides.

Up until about $10^6$ seconds, the largest disturbance is brownian motion of the proof mass due to residual gas collisions which is discussed in Appendix D. There are many other sources of noise in the system, but residual collisions is the only one that results in second-order random walk, i.e. that grows as $t^{3/2}$ not $t^{1/2}$. Furthermore almost all other sources of noise involve rotational and not translation degrees of freedom. After about $10^6$ seconds, the largest non-random disturbances at ambient temperature are electric charge on the proof masses and the radiometer effect. The non-random disturbances are discussed in more detail in Appendix E.

Any satellite-fixed disturbances such as transcollimator light pressure, temperature effects, satellite magnetic fields etc. are roll averaged. Roll-averaging errors essentially arise from two sources: locally-level-fixed miscentering in the cavity and failure to recover from the error induced in the first half roll cycle. For gaps of the order of a cm, if the roll-frequency component of drag-free control error is restricted to $10^{-8}$ meters, the locally-level-miscentering error, $A$, will not exceed $10^{-6}$. For the first half cycle, when a disturbing acceleration is integrated to get the velocity, the integral of one of the roll averaged components is proportional to $1 - c\omega_z t$ which has an average value of one not zero. The position integral is then $\omega_z t - S\omega_z t$ which has a secular growth of $\omega_z t$; and to dominate this error, the number of revolutions must exceed $1 / \pi A$. This requires that the roll period be shorter than $\pi A t$, or about 3 seconds for a $10^6$-second experiment and $A = 10^{-6}$.

## Section 3.4 Long-Term Temperature Drift

Since the proof masses are isolated in a drag-free satellite and since most disturbance forces depend very weakly on the temperature, the primary effects of long term temperature drift are on the readout zero points although it also effects the mass properties and in particular the gravity-gradient cancellation error. Zero-point shift will not be a serious problem, however, because satellite spin separates the experiment data in frequency from the readout errors. The shift in mass properties can be kept in a reasonable bound because the satellite can be made from materials of low thermal coefficient of expansion; because the experiment will be in a thermally insulated enclosure with active temperature control to about 1 K; and because an automatic mass trim system can adjust the satellite center of mass, principal axes of inertia, and gravity-gradient cancellation to compensate for mass distribution changes with temperature.

The satellite will spin while an EP violation will cause a response vector which is parallel to the orbit radius vector. As seen in the satellite, this vector will appear to rotate the rate $-(\omega_z - n)$ where $\omega_z$ is the satellite spin rate and $n$ is the mean orbit rate. The effects of long term temperature drift on the readout then are fixed in the satellite and appear in the signals as slowly varying DC terms, but the experiment data varies at just under spin frequency. This is not roll averaging but frequency separation of the data. All errors such as thermal bending, thermal snap on entering or exiting eclipse (given that the speed of the snap is not too great, if so data can be blanked at this point), systematic variations due to light source aging, detector drift, optical darkening, mechanical distortion, other zero-point drifts, etc. are separated in frequency from the signal due to an EP violation. In other words the thermal system must be constant over one spin period not for an entire experimental run. Temperature variations at roll rate can be reduced to about $10^{-9}$ K by placing the experiment in an insulating chamber with a satellite spin period of a few seconds [15].

Center-of-mass variations perpendicular to the $z$-axis would not exceed $10^{-7}$ meters with a structure of Invar or SuperInvar ($\alpha_{In, \, SIn} \approx 5 \times 10^{-7}$/K, $10^{-7}$/K) and temperature control to 1 K, and the automatic mass-trim system can reduce this to about $10^{-9}$ meters. The principal axis would not rotate more than about $5 \times 10^{-7}$ radians which would not exceed the required spin-axis alignment.

## Section 3.5 Experiment Accuracy: the Kalman-Filter Covariance Results

In order to evaluate the accuracy of the equivalence-principle experiment in the presence of transcollimator noise and gas brownian motion (the dominant disturbance), the covariance equations of a Kalman filter can be solved using the power spectral density of the residual gas collisions as the input noise and the noise equivalent translation of the transcollimator as the measurement noise. A sixth order model including satellite roll and imperfect cancellation is used which estimates the four position and velocity states in the $x$-$y$-plane, $f_x$, and the bias in the $y$-measurement, $b_y$. It is not possible to estimate both $x$ and its bias $b_x$, but this does not prevent $f_x$ from being separately observable. Two cases are shown in Figures 2 and 3 corresponding to different temperatures and pressures and with a readout noise range of to $10^{-12}$ to $10^{-9}$ meters / Hz$^{1/2}$ which can be varied by varying the light-beam power in the transcollimators. The readout noise and the corresponding beam power are marked next to each curve, and it can be seen that the performance is roughly independent of readout noise because a larger readout noise only delays the time before the ultimate accuracy is reached. The noisy curve at the bottom of both figures is a check on the accuracy of the Kalman covariance calculation using a Monte-Carlo simulation of the error in the estimate with a gaussian random number generator and the unknown $x$-bias as an input into the error equation. The noise curve is shown only for the lowest transcollimator readout noise of $10^{-12}$ meters/Hz$^{1/2}$ to avoid cluttering the figures.

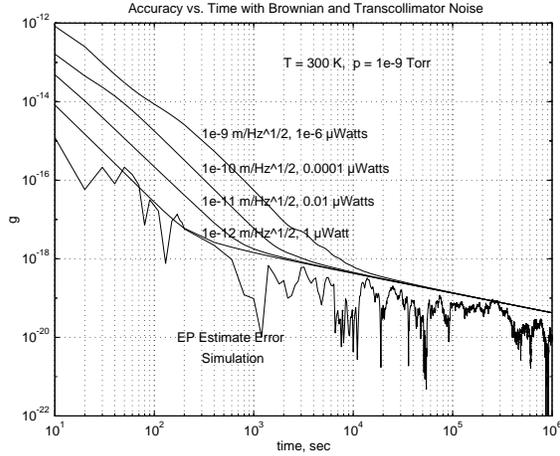 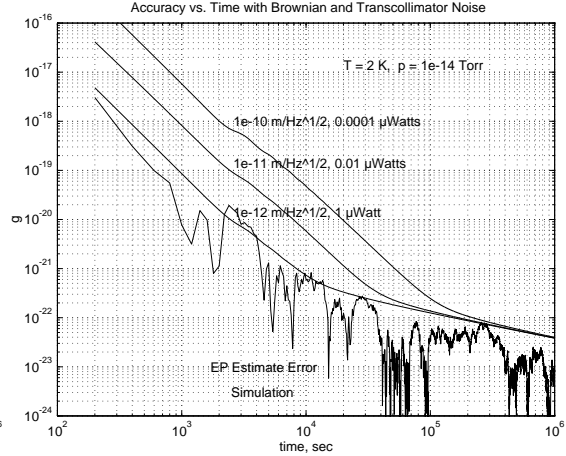

Figure 2.  300 K, $10^{-9}$ Torr, $k^2/3\,n^2 = \pm 10^{-4}$  Figure 3.  2 K, $10^{-14}$ Torr, $k^2/3\,n^2 = \pm 10^{-4}$

At ambient temperatures an accuracy of $4 \times 10^{-19}$ $g$ can be obtained after $10^4$ seconds with any reasonable readout noise, and $4 \times 10^{-20}$ $g$ is possible in an experiment which runs for $10^6$ seconds or about 12 days.  For run times longer than the times given by the cancellation accuracy in Figure C1, the experiment can be restarted and rerun many times during the total life of the satellite.  Since averaging also reduces the error proportional to $t^{1/2}$, the noise performance of a single run of $10^8$ seconds is conceptually equivalent to a series of 100 runs of $10^6$ seconds.  When Figure 2 is extrapolated to three years it shows that $4 \times 10^{-21}$ $g$ is possible at ambient temperatures in an experiment which runs for this term.

In order to do any better than this, however, the vacuum must be greatly improved; and this requires that the experiment must be cooled.  An experiment which ran at 2 K and $10^{-14}$ Torr with liquid Helium could achieve $2 \times 10^{-22}$ $g$ in $3 \times 10^4$ seconds if the heating from a one-microwatt light beam could be tolerated, but it would probably be necessary to use the 0.0001 microwatt beam and wait about $3 \times 10^5$ seconds to get about $6 \times 10^{-23}$ $g$. $10^{-10}$ watts is about 25 times smaller that the radiative cooling shown in Table D1 between a 2 K proof mass and 1.8 K walls.  After 3 years, $4 \times 10^{-24}$ $g$ could be obtained even with a transcollimator noise of $10^{-10}$ meters/Hz$^{1/2}$.

Conversely although an increase in performance requires radical reduction of the pressure, a partial loss of vacuum would not have a great effect on the performance.  For example if a vacuum of $10^{-9}$ Torr cannot be maintained between the shell and the sphere, the performance shown in Figure 2 would worsen as $p^{1/2}$, i.e. for $p = 10^{-7}$ Torr, the accuracy would be 10 times less and for $10^{-5}$ Torr, it would be reduced for example by a factor of only 100 to $10^{-17}$ $g$ after $10^5$ seconds.

The Kalman filter results can be understood intuitively from Figure 4 which shows the noise related items that determine the performance of this experiment.  The set of four solid lines trending downward with slope one-half represent the readout noise from the two-color transcollimators after averaging for the time indicated on the abscissa.  These curves are labeled with the power in the light beams in microwatts because the number of photons per second is important in the design and because heating of the proof masses at cryogenic temperatures is an important problem. The noise equivalent translation for each indicated beam power can be read out directly from the intersection of the curve with the *y*-axis since the origin is at one second. The curves fall slightly below their actual noise equivalent translation because the power spectrum is one-sided so that the averaged noise is reduced by a factor of the square root of two. For example, one $\mu$watt has a noise equivalent translation of $10^{-12}$ meters/Hz$^{1/2}$, etc. Since the photon or shot noise varies as the square root of the number of photons per second, there can be very large reductions in light beam power without greatly affecting the readout noise. Conversely to exceed $10^{-13}$ meters/Hz$^{1/2}$ would require excessively large beam powers even at room temperature. Luckily such extreme performance is not required because as is shown in Figures 2, 3, and 4; the experiment is limited by the disturbances and not the readout noise.

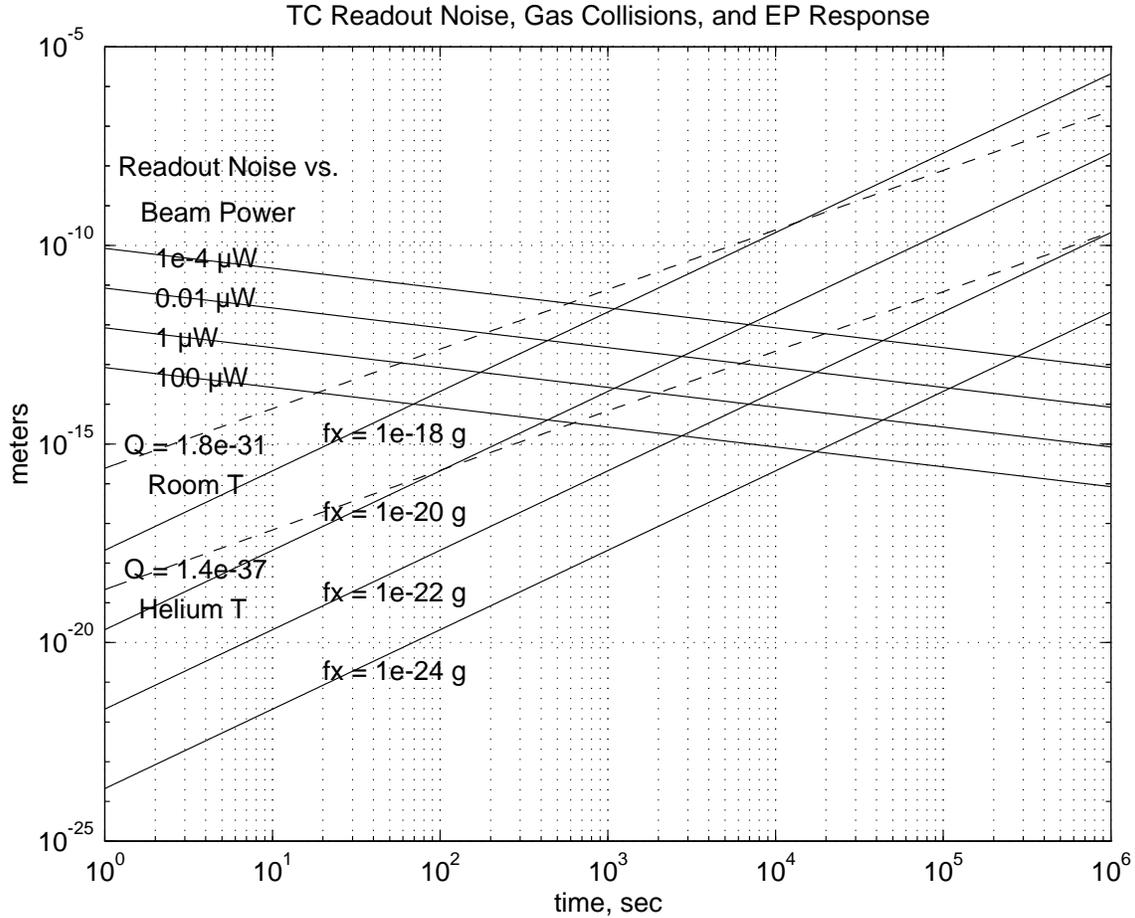

Figure 4. Readout- and Disturbance-Noise Tradeoffs

The solid lines rising from left to right with slope 2 are the response of the canceled system to a violation of the equivalence principle. Figure 4 should be viewed in conjunction with Figure C1 since extending the $t^2$ behavior all the way to $10^6$ seconds assumes that the gravity gradient has been canceled to roughly one part in $10^6$. The validity of the $t^2$ curves can be read off from Figure C1 given an assumption about the accuracy of the cancellation. In reality, however, the results of Figure 4 hold beyond what would be indicated from Figure C1 (see the second paragraph below).

The dashed lines which rise with slope 1.5 represent the random walk of a sphere being bombarded by gas molecules with the indicated power spectral density in $(\text{meters}^2/\text{sec}^4)/\text{Hz}$. Taking as an example the values calculated in Appendix D, for $Q_g = 1.8 \times 10^{-31}$ $(\text{meters}^2/\text{sec}^4)/\text{Hz}$ which corresponds to 300 K and $10^{-9}$ Torr, the response for $10^{-19}$ $g$ comes close enough to the disturbance curve to be detected after about $10^5$ seconds. It can be seen that the readout noise plays no role in this detection as even the curve with the lowest beam power is far below the levels which must be measured. On the other hand detecting $10^{-22}$ $g$ after $10^5$ seconds would require the very low pressures from Helium temperatures. A transcollimator beam power of $10^{-10}$ watt would be able to detect this level, so that heating from the transcollimators is not a problem for these temperatures. At $10^6$ seconds any reasonable beam power is enough to detect $10^{-25}$ $g$.

By comparing Figures 2 and 3 with the curves in Figure 4, the results of the Kalman-Filter covariance calculations in can be better understood. For example, it can be seen that for early times the error in the estimation of $f_x$ descends in the curves in Figures 2 and 3 with a slope of –2.5, and then breaks and continues its descent with a slope of –0.5. The slope of –2.5 indicates that the error is initially dominated by the transcollimator readout noise as is shown in Figure 4. The difference between the time average of the readout noise which descends with slope –0.5 and the response to an EP violation which grows as $t^2$ is $t^{5/2}$. When the gas collisions begin to dominate, the Kalman filter curves break with a new slope of –0.5. The improvement as $t^{1/2}$ is not the result of noise averaging, as can be seen by the fact that the readout noise curves generally lie below the response and disturbance curves; but rather it is due to the fact that the response and disturbance are growing as $t^2$ and $t^{3/2}$ respectively with a difference of $t^{1/2}$. It might be thought that the $t^{1/2}$ improvement would cease as soon as the $k^2$ cancellation error forced the response to leave the $t^2$ curve. It should not be forgotten, however, that the dynamics of the random walk also changes; and apparently the difference still continues to grow as $t^{1/2}$. Thus while the value of $k^2/3\,n^2$ determines the length of time during which the $t^2$ behavior holds, it does not greatly affect the accuracy of a Kalman filter modeled with only brownian disturbance noise and transcollimator readout noise. $k^2/3\,n^2$ is important, however, in that a small value allows the $t^2$ response to last long enough for the signal to grow to a large value even with a small violation of the equivalence principle as is shown in Figure 4. The slight bulge in the part of the curves of Figure 3 descending with slope –2.5 is caused by the cosine term in Equation 9 and can also be seen in Figure C1 a little after 2000 seconds. The slight variations of the curves of Figure 2 between 10 and 100 seconds occur during the transient phase of the filter before the position and velocity errors have settled out. If about a factor of three to ten improvement due to averaging is assumed, then all of the curves in Figures 2 and 3 can be obtained from Figure 4 without the Kalman-filter calculations by matching the values of specific force with which the $t^2$-response curves meet either the transcollimator noise curves or the response curves due to the gas collisions at the various times.

## Section 3.6 Initial-Condition and Bias Errors

Table 3 shows the balance of the errors from the Kalman covariance calculations at $10^6$ seconds. All states except $f_x$, $y$, and $b_y$ reach a steady-state value within a few tens of seconds which is determined by the balance between the transcollimator readout noise and the process noise from the gas collisions, and only $f_x$ and $b_y$ are still being reduced after $10^6$ seconds. It can be seen that the initial conditions can be very quickly measured to high accuracy, and this information can be used to help correct any errors in setting them in later experimental runs.

| | 300 K, $10^{-9}$ Torr, $k^2/3\,n^2 = 10^{-4}$ | | | | 2 K, $10^{-14}$ Torr, $k^2/3\,n^2 = 10^{-4}$ | | |
|---|---|---|---|---|---|---|---|
| RO Noise, m/Hz$^{1/2}$ | $10^{-9}$ | $10^{-10}$ | $10^{-11}$ | $10^{-12}$ | $10^{-10}$ | $10^{-11}$ | $10^{-12}$ |
| $\sigma_x$, m | $3 \times 10^{-11}$ | $5 \times 10^{-12}$ | $1 \times 10^{-12}$ | $2 \times 10^{-13}$ | $8 \times 10^{-13}$ | $2 \times 10^{-13}$ | $3 \times 10^{-14}$ |
| $\sigma_{\dot{x}}$, m/sec | $3 \times 10^{-14}$ | $1 \times 10^{-14}$ | $7 \times 10^{-15}$ | $4 \times 10^{-15}$ | $3 \times 10^{-16}$ | $1 \times 10^{-16}$ | $3 \times 10^{-17}$ |
| $\sigma_y$, m | $1 \times 10^{-11}$ | $4 \times 10^{-12}$ | $1 \times 10^{-12}$ | $2 \times 10^{-13}$ | $1 \times 10^{-13}$ | $4 \times 10^{-14}$ | $1 \times 10^{-14}$ |
| $\sigma_{\dot{y}}$, m/sec | $3 \times 10^{-14}$ | $1 \times 10^{-14}$ | $6 \times 10^{-15}$ | $4 \times 10^{-15}$ | $4 \times 10^{-16}$ | $1 \times 10^{-16}$ | $3 \times 10^{-17}$ |
| $\sigma_{b_x}$, m | $7 \times 10^{-13}$ | $2 \times 10^{-13}$ | $1 \times 10^{-13}$ | $1 \times 10^{-13}$ | $1 \times 10^{-13}$ | $1 \times 10^{-14}$ | $1 \times 10^{-15}$ |

Table 3. Kalman-Filter Covariance Errors

Since they are estimated by the Kalman filter and since they produce no terms which grow as $t^2$, initial-condition errors do not directly enter into the final experiment error. In a long experiment run, however, secular terms from initial conditions can cause the separation between the proof masses to exceed the desirable limit. There are no secular terms in the $y$-axis from initial conditions, and the secular terms in the in the $x$-axis can be seen from Equation B4 to be

$$x = \tfrac{3}{7}\dot{x}_0 t - \tfrac{6}{7} y_0 n t.$$

If it is desired to limit the deviation in a $10^5$-second run to $10^{-8}$ meters, $\dot{x}_0$ must be held to less than $10^{-13}$ m/sec and $y_0$ must be less than about $10^{-10}$ meters. Except for $k^2$, $x_0$ gives no secular contribution. Thus its bias is not as important as $b_y$ which can be estimated by the Kalman filter.

An initial velocity error could be imparted to the shell when it is released at the beginning of an experiment. This error will be less than the control force times the difference in the switching times for opposing forcing electrodes. The switching times can be held equal to better than one microsecond; and since the control specific force will not be larger than about $10^{-11}$ m/sec$^2$, the initial velocity imparted by the switchoff will be less than $10^{-17}$ m/sec which is smaller than the best velocity covariance in Table 3.

**Section 3.7 Control of the Unstable z-Axis**

The $z$-axis is unstable with the gravity-gradient cancellation of Figure 1 with a time constant of approximately 450 seconds. Since $z$ is perpendicular to the experiment plane, however, the problem is manageable; and the best way to deal with it is to actively control the $z$-difference between the two proof masses using the light pressure from the transcollimators. The specific force on the inner sphere from the $z$-components of the

Revision 3.0

beams is about $2P/mc$ where $P$ is the beam power and $m$ is the sphere's mass. For one microwatt in the beam the specific force is about $2 \times 10^{-14}$ m/sec$^2$. This must balance a specific force equal to $5n^2z$ so that if the absolute value of $z$ is less than about $4 \times 10^{-9}$ meters, there is sufficient light pressure from the opposing components to control the differential position of the proof masses. This range can be located prior to the start of the experiment by controlling $z$ with the drag-free controller and the electric fields on the outer sphere and testing until the correct setting to be in range is determined. Since $z$ can be measured by the transcollimators of Figure 1, the only remaining question is the disturbance to the equivalence-principle measurement by the $z$-control force. Assuming that spacecraft roll can be controlled to be perpendicular to the orbit plane to an accuracy of about $10^{-6}$ radians, the disturbance in the $x$-$z$-plane would be $2 \times 10^{-20}$ m/sec$^2 \approx 2 \times 10^{-21}$ $g$ during the time that maximum control was acting. Alternately as is shown in Figure 4, the beam power of the transcollimators can be reduced without significantly increasing the readout noise. Since the experiment is limited by the disturbances and not by the measurement noise, this is also a possible way to keep the $x$-$y$ forces in bound at the price of a tighter range for $z$. Even when low beam power is used for most of the experiment, the beam power can be temporarily turned up for a short time to recenter $z$ and locate the balance point without having a large effect on the results. Thus the unstable $z$-axis can be stabilized using the force from the transcollimator light pressure without inducing excessive disturbances in the $x$-$y$-experiment plane.

## Summary

In the ideal case when a gravity-gradient tensor of the form $\mathrm{diag}\begin{bmatrix} -3n^2 & -3n^2 & 6n^2 \end{bmatrix}$ is generated at the proof masses in a two-sphere drag-free satellite using cancellation masses as shown in Figure 1, and when there are no disturbances; the classical Euler-Hill equations of relative orbital motion are transformed as

$$\begin{array}{ll} \ddot{x} - 3n^2x - 2n\dot{y} = f_x & \ddot{x} - 2n\dot{y} = f_x \\ 2n\dot{x} + \ddot{y} = 0 & \xrightarrow{\quad DC\ Cancellation \quad} \quad 2n\dot{x} + \ddot{y} + 3n^2y = 0. \end{array}$$

The solutions with an initial $x_0$ and an EP violation, $f_x$, are then transformed as

$$\begin{array}{ll} x = x_0 + \left(f_x + 3n^2x_0\right)(1 - cnt)/n^2 & x = x_0 + \dfrac{3}{14}f_xt^2 + \dfrac{4f_x}{49n^2}\left(1 - c\sqrt{7}nt\right) \\ y = -2\left(f_x + 3n^2x_0\right)(nt - Snt)/n^2 & \xrightarrow{DC\ Cancellation} \quad y = -\dfrac{2f_x}{7n}t + \dfrac{2f_x}{7\sqrt{7n^2}}S\sqrt{7}nt \end{array}.$$

The solution of the original Euler-Hill equations shows the two main problems that have prevented a satellite free-fall EP experiment for the last 25 years. The term $f_x + 3n^2x_0$ in the $y$-solution shows for example that an error in the $x$-direction of only $10^{-14}$ meters would mimic an EP violation, $\Delta a/a$, of $10^{-20}$ $g$. The term $nt$ shows that the free-fall response to an EP violation only grows as $t$ not $t^2$.

The solution to the transformed equations shows that in the ideal case, the $3n^2 x_0$ term no longer appears with $f_x$; and the $\frac{3}{14} f_x t^2$ term shows that the DC-canceled EP-violation response grows as $t^2$.

In the practical case where there is a cancellation error, $k^2$, the combination $f_x + k^2 x_0$ replaces $f_x$ in the transformed solutions; and the $t^2$ response lasts between $10^4$ and $10^6$ seconds depending on $k^2$. The term $f_x + k^2 x_0$ means that the non-observability problem is suppressed by the ratio, $k^2 / 3n^2$.

The change in the differential equations of relative motion shows the effect of DC cancellation on the orbit dynamics of a drag-free satellite.

A satellite EP experiment using DC cancellation at ambient temperatures can achieve an accuracy of $10^{-19}$ $g$ after $10^5$ seconds and $4 \times 10^{-20}$ $g$ after a measurement time of $10^6$ seconds (~ 12 days). An experiment which was repeatedly run for 3 years, i.e. $10^8$ seconds, could detect a violation of the equivalence principle as small as $4 \times 10^{-21}$ $g$. This performance can be achieved with reasonable values of the cancellation, $k^2 / 3 n^2 = 10^{-4}$ to $10^{-6}$, and of the readout noise, $10^{-10}$ meters/Hz$^{1/2}$ or larger. The performance is limited by the disturbances not by the transcollimator noise, and the principal disturbance comes from brownian motion of the proof masses caused by random collisions from the residual gas in the cavities. A vacuum of $10^{-9}$ Torr is necessary to get $4 \times 10^{-20}$ $g$ in $10^6$ seconds and no further improvement in the experiment is possible without drastically improving the vacuum, i.e. going to liquid Helium temperatures. At 2 K and $10^{-14}$ Torr it is possible to detect $4 \times 10^{-23}$ $g$ in $10^6$ seconds.

If a Helium-temperature system is compared with a Helium-temperature system, free-falling test masses with DC cancellation can give an improvement of about four orders of magnitude over constrained cylindrical systems [6, 7]; but even an ambient-temperature experiment can give an improvement of about a factor of three. This can be accomplished without difficult cancellation requirements, but $10^{-9}$ Torr may be a problem. Since loss of vacuum only worsens performance as $p^{1/2}$, however, even $10^{-7}$ Torr would only worsen the performance by a factor of 10.

A final important point concerns the ability to compare more than just two materials. A significant criticism of a two-mass experiment is that it can't do this. A multiple-mass comparison must ultimately be performed, but it would be a mistake to assume that this can only be done in a single experiment with multiple sets of differential accelerometers. Experience with prior satellite programs, especially GP-B and Globalstar, suggests that it may in fact be cheaper and easier to fly multiple satellites than to put everything into a single flight. Furthermore an experiment on the Space Station or the Shuttle which launched the experiment overboard in the manner of the Spas series of satellites could repeat the experiment many times with different proof-mass materials, and human assembly of the shell would solve the problem of high-precision assembly of that component in orbit. It should be emphasized, however, that the error analysis for the highest accuracies was done for a low equatorial orbit with the altitude variation limited to a few tens of meters, $A_{att} = 10^{-6}$, geomagnetic shielding, and avoidance of the SAA.

## Appendix A. Equations of Relative Motion of Two Point Masses which Violate the Equivalence Principle

The purpose of this appendix will be to derive the differential equations for the relative motion between two point masses[3] which are not attracted identically by the earth's gravity. It will be shown that, to an excellent approximation, the differential equations of relative motion are just the well known linearized orbit equations driven by a constant specific force equal to the difference in the earth's specific attraction on the two bodies.

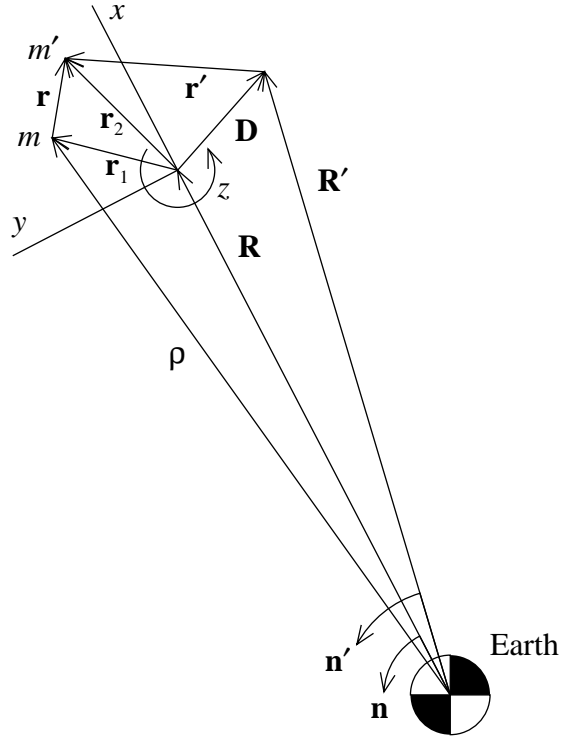

Figure A1. Coordinates for a Free-Fall Equivalence-Principle Experiment

Given two point masses in orbit which violate the equivalence principle, there are two ways to describe their gravitational interaction with the earth which are essentially the same. The bodies may be assumed to have different inertial and gravitational masses, $m_I$ and $m_G$, or it may be assumed that the earth's gravitation constant, $\mu = G M_e$, is different for the two bodies. These two descriptions may be seen to be the same by defining $G_0$ to be a gravitational constant independent of the masses of the bodies and writing $G = G_0 (m_G / m_I)$ and $G' = G_0 (m'_G / m'_I)$ with corresponding definitions for $\mu$ and $\mu'$. For the purposes of this article it will be assumed that with the above exception Newtonian gravity is an adequate approximation. The equations of motion of the two test masses will be linearized about nominal orbits with radius vectors $\mathbf{R}$ and $\mathbf{R}'$ as shown in Figure A1. The equations of motions of the nominal orbits are $\ddot{\mathbf{R}} = -\mu \mathbf{R} / R^3$ and $\ddot{\mathbf{R}}' = -\mu' \mathbf{R}' / R'^3$ where $\mu'$ in the second equation denotes the presence of a small violation of the equivalence principle, i.e. $\Delta\mu \equiv \mu' - \mu \approx 10^{-12}$ to $10^{-24} \times \mu$.

---

[3] The self-gravitation attraction is neglected because point masses are only an approximation to the true physical situation which is a spinning sphere surrounded by a spinning spherical shell. If the shell were perfect, the self attraction would be zero. In the real case, small surface and density variations cause a disturbance (see Table E1).

The equation of the unprimed mass is $\ddot{\rho} = -\mu\rho / \rho^3 + \mathbf{f}_{ns} + \mathbf{f}_d$ with a similar equation for the primed mass. $\mathbf{f}_{ns}$ is the non-spherical part of the gravitational field such as the oblateness terms, etc. and $\mathbf{f}_d$ is the disturbances. $m$ is linearized around the point indicated by the vector, $\mathbf{R}$, which rotates at mean rate, $\mathbf{n}$, given by $n^2 = \mu / R^3$. $m'$ is linearized about the point indicated by the vector, $\mathbf{R}'$; and its location relative to this point is given by $\mathbf{r}'$. $\mathbf{R}'$ rotates at mean rate, $\mathbf{n}'$, given by $n'^2 = \mu' / R'^3$. It is convenient to have two vectors, $\mathbf{r}_2$ and $\mathbf{r}'$, in order to correctly linearize the primed mass about its equilibrium point and at the same time compare it with the unprimed mass from the origin, $\mathbf{R}$. $\mathbf{D}$ is the vector difference between the two equilibrium points such that $\mathbf{R}' = \mathbf{R} + \mathbf{D}$, and the linearized equation for $\mathbf{D}$ is given by

$$\ddot{\mathbf{D}} = -\frac{\mu' \mathbf{R}'}{R'^3} + \frac{\mu \mathbf{R}}{R^3} = -n^2(\mathbf{1} - 3\hat{\mathbf{R}}\hat{\mathbf{R}}) \cdot \mathbf{D} - \frac{\Delta\mu \mathbf{R}}{R^3} - \frac{\Delta\mu}{R^3}(\mathbf{1} - 3\hat{\mathbf{R}}\hat{\mathbf{R}}) \cdot \mathbf{D}. \qquad (A1)$$

where $\hat{\mathbf{R}}$ is the unit $\mathbf{R}$ vector and $\mathbf{1}$ is the unit dyadic, i.e. the unit matrix in vector notation. The right-most term in Equation A1 is smaller than $-n^2(\mathbf{1} - 3\hat{\mathbf{R}}\hat{\mathbf{R}}) \cdot \mathbf{D}$ by the level of the violation of the equivalence principle, $\Delta\mu / \mu$, and can be safely neglected. When this term is omitted, Equation A1 becomes

$$\ddot{\mathbf{D}} = -n^2(\mathbf{1} - 3\hat{\mathbf{R}}\hat{\mathbf{R}}) \cdot \mathbf{D} - \frac{\Delta\mu \mathbf{R}}{R^3}. \qquad (A2)$$

Except for a small practical problem, if $m$ and $m'$ were actually located at $\mathbf{R}$ and $\mathbf{R}'$, Equation A2 with the disturbances included would be the desired differential equation for the relative motion; and no further derivation would be needed. Equation A2 is the well-known linearized orbit equation driven by a small term which is equal to the difference in the earth's specific attraction of the two masses, i.e. by the violation of the equivalence principle. The only assumptions involved in Equation A2 are linearity and the neglect of the right-most term of Equation A1. The practical problem is that Equation A2 is most conveniently solved if it results in linear differential equations with constant coefficients when it is put in coordinate form. In order for this to occur, the nominal equilibrium point, $\mathbf{R}$, must be in a circular orbit so that the rotation rate, $\mathbf{n}$, of the x-y-axes is constant and the unit vector, $\hat{\mathbf{R}}$, does not change direction relative to the mean rate as would be the case with an elliptical orbit. If $m$ and $m'$ are not located at $\mathbf{R}$ and $\mathbf{R}'$, then $\mathbf{R}$ and $\mathbf{R}'$ can be assumed to describe circular orbits; and practical aspects of real orbits such as ellipticity and perturbations such as $J_2$ can be taken into account by having the actual orbit of the test masses remain within a few kilometers of the nominal circular orbit. In this case small ellipticities can be accounted for by the initial conditions; and $J_2$, etc. can be driving terms accounted for in $\mathbf{f}_{ns}$ and $\mathbf{f}'_{ns}$. When this is done, the differential equation for the relative motion remains the same as Equation A2; but one must examine the assumptions for its validity a little more carefully. The equation for relative motion will now be derived for the case where the test masses are close to the equilibrium orbits which will be assumed to be circular once the actual components of the equations are written.

Neglecting the disturbances for the moment, when $\rho'$ (not shown in the figure) is linearized about $\mathbf{R}'$, the standard linear equation of motion of $\mathbf{r}'$ is obtained

$$\ddot{\mathbf{r}}' = -n'^2(\mathbf{1} - 3\hat{\mathbf{R}}'\hat{\mathbf{R}}') \cdot \mathbf{r}' \qquad (A3)$$

Since $\mathbf{r}_2 = \mathbf{r}' + \mathbf{D}$, Equations A1 and A3 may be added to obtain the equation for $\mathbf{r}_2$,

$$\ddot{\mathbf{r}}_2 = -n^2(\mathbf{1} - 3\hat{\mathbf{R}}\hat{\mathbf{R}}) \cdot \mathbf{r}_2 - \frac{\Delta\mu\mathbf{R}}{R^3} - \frac{\Delta\mu}{R^3}(\mathbf{1} - 3\hat{\mathbf{R}}\hat{\mathbf{R}}) \cdot \mathbf{D}$$

$$-[n'^2(\mathbf{1} - 3\hat{\mathbf{R}}'\hat{\mathbf{R}}') - n^2(\mathbf{1} - 3\hat{\mathbf{R}}\hat{\mathbf{R}})] \cdot (\mathbf{r}_2 - \mathbf{D}). \qquad (A4)$$

At this point it should be emphasized that no other approximations have been made in deriving Equations A1, A3, and A4 except linearization; that is, the squares and higher order terms in $\mathbf{r}'$, $\mathbf{D}$, and $\mathbf{r}_2$ divided by either $R$ or $R'$ have been neglected. The right-most two terms of Equation A4 are linear, but they also involve $\Delta\mu/\mu$ and are therefore much smaller than the corresponding term just to the right of the equal sign in Equation A4. When the last two terms in Equation A4 are neglected, the equation for the motion of the primed mass relative to the equilibrium point of the unprimed mass, $\mathbf{R}$, becomes

$$\ddot{\mathbf{r}}_2 = -n^2(\mathbf{1} - 3\hat{\mathbf{R}}\hat{\mathbf{R}}) \cdot \mathbf{r}_2 - \frac{\Delta\mu\mathbf{R}}{R^3}. \qquad (A5)$$

If $\mathbf{r}_1$ is defined to be the difference between the unprimed mass and its equilibrium point, $\mathbf{R}$; then the linear equation for $\mathbf{r}_1$ can be shown to be $\ddot{\mathbf{r}}_1 = -n^2(\mathbf{1} - 3\hat{\mathbf{R}}\hat{\mathbf{R}}) \cdot \mathbf{r}_1$ in the same way as was done for Equations A1 and A3. If $\mathbf{r}$ is defined to be the difference between $\mathbf{r}_2$ and $\mathbf{r}_1$, then the equation for $\mathbf{r}$ may be obtained by subtracting the equation for $\mathbf{r}_1$ from the one for $\mathbf{r}_2$

$$\ddot{\mathbf{r}} = -n^2(\mathbf{1} - 3\hat{\mathbf{R}}\hat{\mathbf{R}}) \cdot \mathbf{r} + \mathbf{f} + \mathbf{f}_d + \Delta\mathbf{f}_{ns} \qquad (A6)$$

where $\mathbf{f} \equiv -\Delta\mu\mathbf{R}/R^3$ is the violation of the equivalence principle which is to be measured, $\mathbf{f}_d$ comes from the disturbances, and $\Delta\mathbf{f}_{ns} = \mathbf{f}'_{ns} - \mathbf{f}_{ns}$. Equation A6 is the equation for the difference in the effective "gravitational centers" of the two spherical proof masses in the center of the drag-free satellite cavity and is the equation which was to be derived in this appendix.

The additional approximations beyond linearization used in deriving Equation A6 may be summarized as follows:

1)   The assumed violation of the equivalence principle which is to be searched for:

$$\mu' - \mu \equiv \Delta\mu = (10^{-12} \text{ to } 10^{-24})\mu. \qquad (A7)$$

2) The difference between the equilibrium points of the two nominal orbits:

$$\frac{D}{R} = \frac{10^{-8} \text{ to } 10^{-14} \text{ meters}}{10^7 \text{ meters}} = 10^{-15} \text{ to } 10^{-21}. \tag{A8}$$

3) The difference in the squares of the mean rates of the two equilibrium points:

$$n'^2 - n^2 = \frac{\mu'}{R'^3} - \frac{\mu}{R^3} = (10^{-12} \text{ to } 10^{-24})n^2 - 3n^2 \hat{\mathbf{R}} \cdot \frac{\mathbf{D}}{R} \approx (10^{-12} \text{ to } 10^{-21})n^2. \tag{A9}$$

4) The difference between using the primed or the unprimed reference frame:

$$n'^2 (\mathbf{1} - 3\hat{\mathbf{R}}'\hat{\mathbf{R}}') - n^2 (\mathbf{1} - 3\hat{\mathbf{R}}\hat{\mathbf{R}})$$

$$\approx (10^{-12} \text{ to } 10^{-24})n^2 (\mathbf{1} - 3\hat{\mathbf{R}}'\hat{\mathbf{R}}') - 3n^2 \left[ \hat{\mathbf{R}}\frac{\mathbf{D}}{R} + \frac{\mathbf{D}}{R}\hat{\mathbf{R}} \right]$$

$$\approx (10^{-12} \text{ to } 10^{-21})n^2 \mathbf{1}. \tag{A10}$$

5) The rate at which the primed and unprimed reference frames rotate apart:

$$n' - n = \frac{n'^2 - n^2}{2n} \approx 0.5 \times (10^{-12} \text{ to } 10^{-21})n. \tag{A11}$$

After one year, $\approx 3 \times 10^7$ seconds, the value of $(n'-n)t$ is only about $10^{-8}$ to $10^{-17}$ radians so that any misalignment between the unit vectors, $\hat{\mathbf{R}}$ and $\hat{\mathbf{R}}'$, can safely be neglected. Also notice that although $r_1$ and $r_2$ can be quite large, of the order of kilometers, the ratios $r_1/R$, $r_2/R'$, etc. which can be of the order of $10^{-5}$ to $10^{-6}$ do not occur in the list of approximations necessary to combine Equation A5 and the equation for $\mathbf{r}_1$ into Equation A6. On the other hand since $\mathbf{r}$ and $\mathbf{D}$ are of the order of $10^{-8}$ to $10^{-14}$ meters, Equations A2 and A6 are very accurate. Some care has been exercised in spelling out precisely what has been neglected in deriving Equation A6 because, while linearization is a standard well understood procedure, neglecting a linear term simply because it is small can be dangerous unless it is done by comparison with some much larger identical term.

Up to this point if $n$ and $n'$ are not interpreted to be mean rates, Equations A1 through A6 are valid for any nominal reference orbit, elliptical, circular, or hyperbolic; but the rest of the paper will now assume that the reference orbits are circular. Appendix B will resolve the components of Equations A2 or A6 in the locally-level orbit coordinate system of Figure A1 and, in addition, will include the effects of the gravity gradients of the local masses in the drag-free satellite.

**Appendix B. The Equations of Relative Motion with a Circular Reference Orbit**

**Appendix B1. Locally-Level Equations in a Spinning Drag-Free Satellite Including the Gravity Gradients**

This appendix applies the equations of relative motion of the last section to two concentric spheres in the locally-level orbit reference frame shown in Figure A1 for a spinning drag-free satellite with its spin axis perpendicular to the orbit plane. The components of the difference vector, $\mathbf{r}$, are $x$, $y$, and $z$ in the locally-level frame.

When the gravity gradient terms from the satellite masses are included in Equations A2 or A6, the components of the linearized equations for the relative motion are given by

$$\begin{bmatrix} \ddot{x} - n^2 x - 2n\dot{y} \\ 2n\dot{x} + \ddot{y} - n^2 y \\ \ddot{z} \end{bmatrix} = \left\{ \begin{bmatrix} 2n^2 & 0 & 0 \\ 0 & -n^2 & 0 \\ 0 & 0 & -n^2 \end{bmatrix} + \begin{bmatrix} \frac{1}{2}(g_{11} + g_{22}) & 0 & 0 \\ 0 & \frac{1}{2}(g_{11} + g_{22}) & 0 \\ 0 & 0 & g_{33} \end{bmatrix} + \right.$$

$$\left. \begin{bmatrix} \frac{1}{2}(g_{11} - g_{22})c2 - g_{12}S2 & \frac{1}{2}(g_{11} - g_{22})S2 + g_{12}c2 & g_{13}c - g_{23}S \\ \frac{1}{2}(g_{11} - g_{22})S2 + g_{12}c2 & -\frac{1}{2}(g_{11} - g_{22})c2 + g_{12}S2 & g_{13}S + g_{23}c \\ g_{13}c - g_{23}S & g_{13}S + g_{23}c & 0 \end{bmatrix} \right\} \begin{bmatrix} x \\ y \\ z \end{bmatrix} + \begin{bmatrix} f_x + f_{dx} + \Delta f_{nsx} \\ f_{dy} + \Delta f_{nsy} \\ f_{dz} + \Delta f_{nsz} \end{bmatrix}$$

$$(B1)$$

The terms $n^2 x$ and $n^2 y$ on the left side are from the centrifugal force of the rotating coordinate system and the terms $2n\dot{x}$ and $2n\dot{y}$ are from the coriolis force. On the right side are the gravity-gradient tensor from the earth, $-n^2(\mathbf{1} - 3\hat{\mathbf{R}}\hat{\mathbf{R}})$ and the gravity-gradient tensor from rotating satellite masses. $f_x$ is a constant specific force which is used to model a violation of the equivalence principle as defined in Equation A6, $f_{dxyz}$ represents the disturbance force errors from the satellite, and $\Delta f_{nsxyz}$ the difference in the non-spherical gravitation terms. The gravity gradients of the satellite masses are assumed to be described by a constant matrix in the satellite reference frame, $g_{ij}$. $g_{11}$, $g_{22}$, and $g_{33}$ are assumed to be large, of the order of $3\,n^2 \approx 3 \times 10^{-6}/\mathrm{sec}^2$, while $g_{12}$, $g_{13}$, $g_{23}$, and $g_{11}$ - $g_{22}$ are assumed to be small, of the order of $10^{-10}$ to $10^{-12}/\mathrm{sec}^2$. $S$ and $c$ are the sine and cosine of the satellite rotation angle which is assumed to be spinning about the z-axis, and $S2$ and $c2$ are the sine and cosine of two times the rotation angle. The form of the gravity-gradient tensor for the satellite masses in Equation B1 results from the tensor being transformed from the satellite axes, to the locally-level frame.

The possibility of using special masses to achieve the $g_{ii}$'s of Equation 6 in the main text can be seen as follows: for a spherical mass of radius, $a$, the gravity-gradient tensor at a distance, $d$, from the center is given by

$$T_{gg\,sphere} = \frac{4\pi G\rho}{3}\left(\frac{a}{d}\right)^3 \begin{bmatrix} -1 & 0 & 0 \\ 0 & -1 & 0 \\ 0 & 0 & 2 \end{bmatrix}, \tag{B2}$$

so that at the surface of a sphere the gravity-gradient in the cross direction is simply $4\pi G\rho/3$, and depends only on the density. Equation B2 is written for the case where the $z$-axis is the radial axis of the sphere, and $x$ and $y$ are the cross axes. This corresponds to the spheres in Figure 1. Since the gravity-gradient at the surface depends only on the density it is easy to choose masses which equal or exceed the earth's gravity gradient. The denser the masses, the less must be the total overall mass; and for this reason, heavy metals are chosen, Osmium, Iridium, or Platinum.

If it is assumed that the satellite masses have been trimmed so that the off-diagonal terms can be neglected in Equation B1, Equations 3, 4, and 5 in the main text are obtained.

## Appendix B2. Perfect DC Gravity-Gradient Cancellation

When the masses of Figure 1 are chosen such that $(g_{11} + g_{22})/2 = -3\,n^2$ and $g_{11} - g_{22} = 0$, i.e. $4\pi G\rho a^3/3\,d^3 = 1.5\,n^2$ in Equation B2, the $3\,n^2$ term in Equation 3 is canceled; and the Euler-Hill equations ($\ddot{x} - 3n^2 x - 2n\dot{y} = f_x$ and $2n\dot{x} + \ddot{y} = 0$) are transformed into

$$\ddot{x} - 2n\dot{y} = f_x, \tag{7}$$

$$2n\dot{x} + \ddot{y} + 3n^2 y = 0, \tag{8}$$

and

$$\ddot{z} - 5n^2 z = 0. \tag{B3}$$

The Laplace-transform solution of the $x$ and $y$ equations is

$$\begin{bmatrix} x \\ y \end{bmatrix} = \frac{\begin{bmatrix} s^2 + 3n^2 & 2ns \\ -2ns & s^2 \end{bmatrix}}{s^2(s^2 + 7n^2)}\left[ \begin{pmatrix} f_x/s \\ f_y/s \end{pmatrix} + \begin{pmatrix} sx_0 \\ 2nx_0 \end{pmatrix} + \begin{pmatrix} \dot{x}_0 \\ 0 \end{pmatrix} + \begin{pmatrix} -2ny_0 \\ sy_0 \end{pmatrix} + \begin{pmatrix} 0 \\ \dot{y}_0 \end{pmatrix} \right] \tag{B4}$$

which gives Equations 9 and 10 in the time domain.

## Appendix C. Imperfect Cancellation

In order to determine the effect of imperfect cancellation, it is necessary to solve Equations 3 and 4 for the case where $k^2 \equiv 3\,n^2 + (g_{11} + g_{22})/2$ is not exactly zero. In order to avoid the problem of periodic coefficients, however, it will be assumed that $g_{11} - g_{22}$ is negligibly small. There are two cases which must be separately calculated, $k^2 > 0$ and $k^2 < 0$. In both cases the equations are

$$\ddot{x} - k^2 x - 2n\dot{y} = f_x \tag{C1}$$

$$2n\dot{x} + \ddot{y} + (3n^2 - k^2)y = 0, \tag{C2}$$

but the roots of the characteristic equations are different in the two cases.

## Appendix C1. $k^2$ Greater Than Zero

In this case, there are two real roots and two imaginary roots; and the system is slightly unstable. This is not a serious problem, however, since the unstable time constant varies between $10^4$ and $10^6$ seconds depending on the precision of the cancellation. The instability is actually an advantage since it can be used to measure the accuracy of the cancellation when it is implemented with a mass-trim system of movable masses.

If the two roots are defined to be

$$\omega^2 = \tfrac{7}{2}n^2 - k^2 + K \tag{C3}$$

and

$$a^2 = -\tfrac{7}{2}n^2 + k^2 + K \tag{C4}$$

where $K$ is the radical

$$K = \sqrt{\tfrac{49}{4}n^4 - 4n^2 k^2}, \tag{C5}$$

the exact solutions are

$$x = -\frac{f_x}{k^2} + (f_x + k^2 x_0)\left[ -\left(\frac{1}{2} + \frac{n^2}{4K}\right)\frac{c\omega t}{\omega^2} + \left(\frac{1}{2} - \frac{n^2}{4K}\right)\frac{Cat}{a^2} \right] \tag{C6}$$

and

$$y = -\frac{n(f_x + k^2 x_0)}{K}\left[ \frac{-S\omega t}{\omega} + \frac{Sat}{a} \right] \tag{C7}$$

where $S, c, S,$ and $C$ are the trigonometric and hyperbolic sines and cosines respectively. In the limit as $k^2 \to 0$, i.e. as the accuracy of the cancellation improves, Equations C6 and C7 become

$$x \xrightarrow{k^2 \to 0} x_0 + \frac{3}{14}(f_x + k^2 x_0)t^2 + \frac{4(f_x + k^2 x_0)}{49n^2}(1 - c\sqrt{7}nt) \tag{C8}$$

and

$$y \xrightarrow{k^2 \to 0} -\frac{2}{7}\frac{(f_x + k^2 x_0)}{n}t + \frac{2(f_x + k^2 x_0)}{7\sqrt{7}n^2}S\sqrt{7}nt \tag{C9}$$

which agrees with the Equations 9 and 10 for perfect cancellation with the very important exception that the combination $f_x + k^2 x_0$ always appears together. Thus there is still a radial non-observability problem with this method, but it can be suppressed by the accuracy of cancellation.

## Appendix C2. $k^2$ Less Than Zero

In this case, there are two pairs of imaginary roots; and the system is stable. If the two roots are defined to be

$$\omega_1^2 = \tfrac{7}{2} n^2 - k^2 + K \tag{C10}$$

and

$$\omega_2^2 = \tfrac{7}{2} n^2 - k^2 - K \tag{C11}$$

where $K$ is the same radical as in Equation C5, the exact solutions are

$$x = -\frac{f_x}{k^2} + (f_x + k^2 x_0)\left[ -\left(\frac{1}{2} + \frac{n^2}{4K}\right)\frac{c\omega_1 t}{\omega_1^2} - \left(\frac{1}{2} - \frac{n^2}{4K}\right)\frac{c\omega_2 t}{\omega_2^2} \right] \tag{C12}$$

and

$$y = -\frac{n(f_x + k^2 x_0)}{K}\left[ \frac{S\omega_1 t}{\omega_1} - \frac{S\omega_2 t}{\omega_2} \right]. \tag{C13}$$

In the limit as $k^2 \to 0$ Equations C12 and C13 also go to Equations C8 and C9, so that the cancellation limit is independent of sign.

## Appendix C3. Deviation Due to Imperfect Cancellation

Figure C1 shows the deviation of the response from the ideal $t^2$, i.e. from Equation 9, when the cancellation is not perfect. The dotted line in the lower left corner of the figure and its solid extension to the upper right is the curve $3 f_x t^2 / 14$ with $f_x = 10^{-20} g$. The slightly higher solid line above the dotted portion is the variation due to the term, $4 f_x (1 - c\sqrt{7}nt) / 49 n^2$ in Equations 9 or C8. In the upper right are the curves for the case $k^2 > 0$ which deviate exponentially due to the hyperbolic cosine term in Equation C6, and they are marked with the values of $k^2 / 3 n^2$, the cancellation accuracy. Just below these curves are those for $k^2 < 0$, Equation C12; and they are also marked with the values of $k^2 / 3 n^2$. Their oscillatory shapes are due to the $c\omega_2 t$ term and the distortion of the logarithmic plot. On a normal linear plot these are simply cosine terms oscillating between zero and the maximum value shown for each case. The important result shown in Figure C1 is that if the gravity gradient can be canceled to $10^{-6}$, the $t^2$ behavior holds out to about $10^6$ seconds; and even if the cancellation is as poor as $10^{-2}$, the approximation is still valid to $10^4$ seconds, almost 3 hours. With movable cancellation masses Figure C1 shows that in about $10^4$ to $10^6$ seconds, the large exponential signal makes it possible to calibrate the system to the corresponding value of $k^2/3\,n^2$ shown on the plot by setting the trim masses to slightly excite the unstable root and observing the resulting deviation.

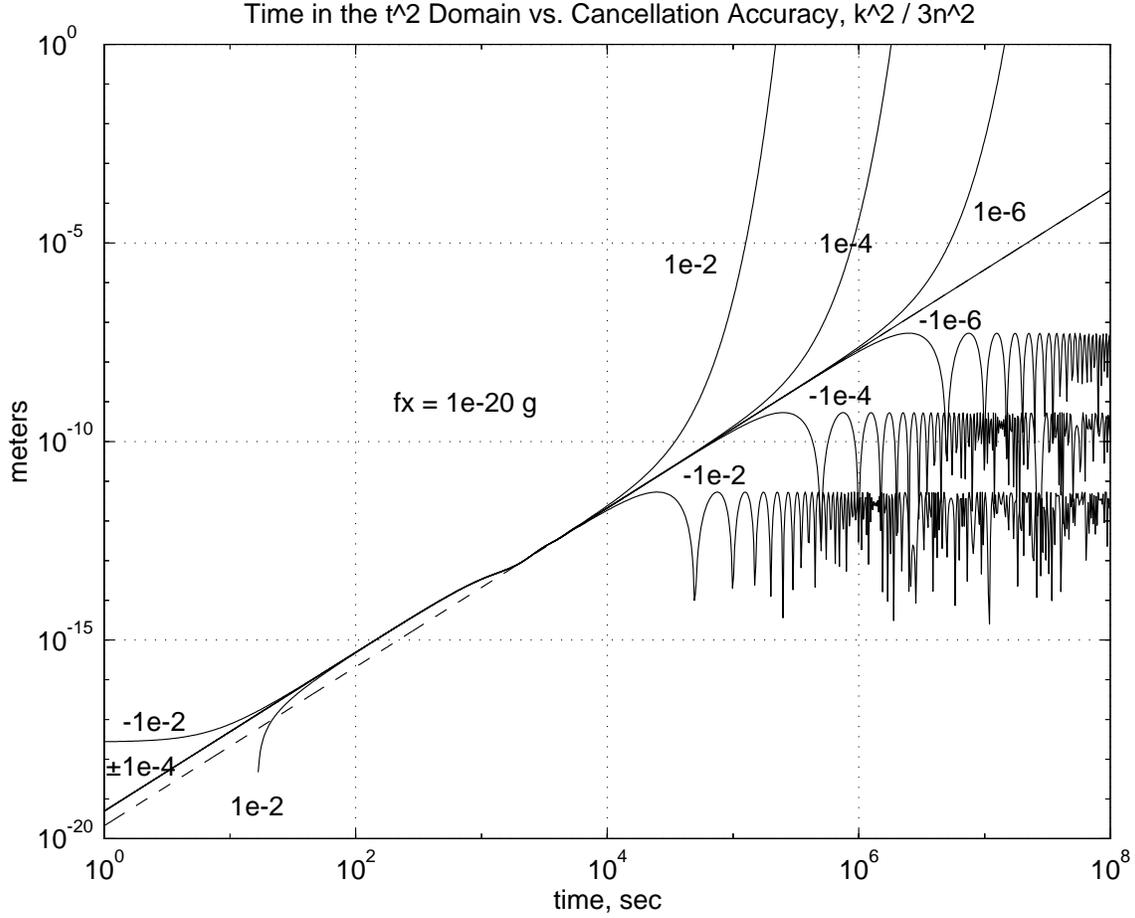

Figure C1. Imperfect Cancellation

## Appendix D. Brownian Motion of the Proof Masses

At ambient temperature with rapid satellite spin and for $t < 10^6$ seconds, the largest disturbance is brownian motion (Figure 4) which causes the two proof masses execute second order random walk as a result of gas collisions. The single-axis random walk of a spherical mass is

$$\langle x^2 \rangle = \frac{2k_BT}{d_g}\left[t - \frac{2m}{d_g}\left(1 - e^{-d_g t/m}\right) + \frac{m}{2d_g}\left(1 - e^{-2d_g t/m}\right)\right] \xrightarrow{t \to 0} \frac{2d_g k_B T}{3m^2}t^3 = \frac{1}{3}Q_g t^3 \quad \text{(D1)}$$

where $m$ is the mass of the sphere or the shell, $d_g$ is the damping of the mass's motion caused by the residual gas in the cavity, and $Q_g$ is the power spectral density from gas collisions in $(\text{meters}^2/\text{sec}^4)/\text{Hz}$. The damping is given by

$$d_g = 3pa^2 \left( \frac{2\pi m_{av}}{k_B T} \right)^{1/2}. \tag{D2}$$

where $p$ is the residual gas pressure in the cavity, $a$ is the radius of the sphere, $m_{av}$ is the average molecular mass of the gas, $k_B$ is the Boltzmann constant, and $T$ is the absolute temperature. The important point of Equation D1 is that for about the first 30,000 years ($\approx m / d_g$), the residual gas causes the proof-mass error to increase as $t^{3/2}$. This means that if the only indication of a violation of the equivalence principle were the drift of Equation 2 proportional to $t$ (the case without gravity-gradient cancellation), the error due to the gas collisions would actually increase with time as $t^{1/2}$. This places special importance on the fact that when the $3n^2$ term is canceled, there is a solution proportional to $f_x$ which grows as $t^2$ with the result that the error with random walk decreases as $t^{1/2}$ instead of growing.

Equation D1 is the gas brownian motion only for a single-axis model, but it is possible to solve the Kalman-filter covariance equation for Equations 3 and 4 using the measurement noise from the transcollimator and the process noise from gas brownian motion. The power spectral density of the brownian noise can be obtained from the equipartion-of-energy theorem and is given by

$$Q_g = \frac{2d_g k_B T}{m^2} = \frac{6pa(2\pi m_{av} k_B T)^{1/2}}{m^2}. \tag{D3}$$

This power spectral density multiplied by three (since three surfaces experience collisions) is used in Section 3.5 in the Kalman-filter covariance calculation.

In addition to residual gas in the proof-mass cavities there is a spectrum of penetrating radiation in orbit which besides heating and charging the proof masses can also cause random walk of the proof masses in the same manner as the gas collisions. For an order-of-magnitude estimate, the damping and power spectral density of the radiation may be calculated from the above expressions for the residual gas by replacing $3k_B T$ with $(\Delta P)^2 / 4 m_r$ where $\Delta P$ is the momentum transferred to the proof mass by the collision and $m_r$ is the mass of the penetrating radiation. With this approximation and the assumption that the radiation is non-relativistic, the damping and power spectral density are given by

$$d_r = \frac{FA\Delta P m_r}{P} \tag{D4}$$

and

$$Q_r = \frac{FA(\Delta P)^3}{6Pm^2} \tag{D5}$$

where $F$ is the particle flux, $A$ is the area of the proof mass, and $P$ is the particle momentum given by $P = \sqrt{2m_r E}$ with $E$ equal to the energy per particle in joules. (The Iron cosmic rays are relativistic so that $P \approx E / c$, but the stopping power has been calculated non-relativistically.) The radiation environment for the STEP experiment is discussed by Jafry and Tranquille [16] which also gives the non-relativistic expression for

Revision 3.0

calculating $\Delta P$. Values for $F$ and $E$ are taken from Figure 1 of that reference and from U.S. Navy Creme96 calculations (http://crsp3.nrl.navy.mil/creme96).

The large-particle calculation is done in a similar manner to the radiation but the results shown here are very conservative since it is expected that collisions will not occur for more than a fraction of a second due to trapping of the particles on the walls. In essence the large particles are swept away by the rotation of the rotors until they are all trapped into what is known in clean-room terminology as the "quiescent state". It is also assumed that particles larger than about one micron will be eliminated by clean-room assembly procedures.

| Residual Gas | T<br>K | p<br>Torr | n<br>$\#/cm^3$ | R<br>watts | $d_g$<br>N/(m/s) | $Q_g$<br>$m^2/s^4/Hz$ |
|---|---|---|---|---|---|---|
| Ambient Temperature | 300 | 1.0E-09 | 3.2E+07 | 4.5E-01 | 5.0E-13 | 1.8E-31 |
| Helium Temperature | 2 | 1.0E-14 | 4.8E+04 | 8.8E-10 | 6.1E-17 | 1.4E-37 |
| | | | | | | |
| Large-Particles | Av Vel<br>m/sec | Collisions<br>#/yr | Radius<br>microns | P<br>kg m/s | $d_c$<br>N/(m/s) | $Q_c$<br>$m^2/s^4/Hz$ |
| Micron-Size | 3.0E-02 | 100 | 0.5 | 3.7E-17 | 7.8E-21 | 1.8E-37 |
| | | | | | | |
| Unshielded Radiation | E<br>mev | F<br>$\#/cm^2/s$ | P<br>kg m/s | $\Delta P$<br>kg m/s | $d_r$<br>N/(m/s) | $Q_r$<br>$m^2/s^4/Hz$ |
| Peak Trapped Electrons | 3.0E-01 | 3.0E+08 | 3.0E-22 | 3.0E-22 | 1.9E-21 | 1.3E-33 |
| Galactic Cosmic Rays | 1.0E+02 | 1.0E+00 | 2.3E-19 | 2.3E-19 | 1.2E-26 | 2.7E-36 |
| Iron Cosmic Rays | 1.0E+06 | 5.0E-03 | 5.3E-16 | 1.6E-18 | 1.8E-31 | 2.1E-39 |
| | | | | | | |
| Heavy Metal Decay<br>Surface layer escapes | E<br>mev | Decay<br>#/sec | Range<br>meter | P<br>kg m/s | $d_d$<br>N/(m/s) | $Q_d$<br>$m^2/s^4/Hz$ |
| Uranium *a* Decay | 5 | 2.9E+06 | 5.35E-05 | 1.0E-19 | 2.1E-23 | 2.4E-34 |

Table D1. Power Spectral Densities of Residual Gas, Large Particles, and Radiation

Table D1 calculates the power spectral density in $(m^2/sec^4)/Hz$ for residual gas at room temperature and at cryogenic temperatures as well as for large-particle collisions and for the radiation environment described in Figure 1 of Reference [16]. It gives the differential motion of a sphere and shell with a mass of 0.26 kg each assuming that both sides of the shell are bombarded, i.e. Equations D3 and D5 are multiplied by three. The radius of the central sphere was taken as 1.5 cm and the radii of the shell as 2.0 and 3.25 cm with densities of 18,950 and 2340 kg/m³ respectively. For the radiation, Table D1 gives the values of $Q_r$ without including the effect of any shielding so that in the real situation they would be much smaller. Trapped protons both in and out of the SAA and solar flares have been looked at but are not included in Table D1 because the both Reference [16] and the Creme96 calculations show that they are not present in a 500-km equatorial orbit. In the case of trapped electrons and the low-energy range of galactic cosmic rays (which has the highest fluxes), $\Delta P$ is assumed to be the same as $P$, i.e. they would be completely stopped by the proof masses. For this reason they will also be shielded by the magnetic and thermal shields and the spacecraft structure and will not cause significant proof-mass random walk; i.e. if they can be stopped by the proof masses, they can be stopped by the ambient shields. In the case of Uranium alpha decay, a low temperature experiment may require that some other non-radioactive heavy metal be used for the central proof mass at some loss in the expected EP-violation signal.

Thus the power spectral density of the residual gas particles dominates the disturbances from particle collisions. Because cooling of the proof masses is difficult at the cryogenic temperatures, the radiative cooling power, $R$, is also included for the residual gas values. The wall temperature is assumed to be 270 K for ambient temperatures and 1.8 K for Helium temperatures. Because the residual gas collisions represent the largest disturbance, the values of $Q_g$ in Table D1 will be used in Section 3.5 to calculate the expected accuracy of the equivalence-principle experiment.

## Appendix E. Disturbances Not Modeled as Random Collisions

It is useful to divide the disturbances into two classes, random and non-random, because a direct comparison of the disturbance in $g$'s is not possible for random collisions. It can be seen from Figure 4, however, that gas collisions dominate any room temperature disturbance less than about $10^{-19}$ $g$ and any low temperature disturbance less than about $10^{-22}$ $g$ for times less than $10^6$ seconds. The calculation of those errors modelled as non-random is guided by the following principle: If the errors are small with respect to the above limits, a rough order of magnitude estimate is sufficient; and if the errors are close to the limits either an exact calculation or an upper bound is needed.

If an error source is exactly parallel to the orbit radius vector, it can mimic an EP violation; and special care must be taken with these errors. Examples (see below) are the dumbbell effect and the Lorentz force. If this is not the case, the calculations of the error forces below overestimate the effect of the error source on the experiment error.

At ambient temperature the largest disturbing force between the two spheres comes from electric charge, both from the Lorentz force and direct electric attraction.[4] Figure E1 shows

---

[4] The electric force from charges on the conducting sphere and shell can be calculated by differentiating the expression for the stored electrostatic energy with respect to the miscentering. The nested system consists of the inner sphere surrounded by the shell which is in turn surrounded by the cavity wall and the satellite. If the charge on the sphere is $q_1$, the shell $q_2$, and the satellite $q_3$; the charges on the sphere and the inner shell surfaces are $q_1$ and $-q_1$, on the outer shell surface and the cavity surface $q_1 + q_2$ and $-(q_1 + q_2)$, and on the satellite surface $q_1 + q_2 + q_3$. By bringing a small test charge in from infinity, it can be shown that elastances are given by $s_{13} = s_{23} = s_{31} = s_{32} = s_{33}$ and $s_{12} = s_{21} = s_{22}$. The form of the capacitance matrix can be found by inverting the matrix of elastances; and by inverting the capacitance matrix again, it can be shown that

the specific force in $g$'s acting on the sphere or the shell due to miscentering of the surfaces carrying an electric charge. From the charge levels in the figure it can be seen that it will be necessary to measure the charge on the sphere and shell and perhaps to actively discharge them.[5] In addition to charge, Figure E1 also shows the disturbance from the amplitude of Helium tides assumed for MiniSTEP [6]. The electric disturbance from miscentering

$s_{11} = 1/c_{11} - 1/c_{23} + 1/C_3$, $s_{22} = -1/c_{23} + 1/C_3$, and $s_{33} = 1/C_3$ where $c_{11} = c_{11S}(a, b, x_{ce})$, $c_{23} = -c_{11S}(c, d, x_{ce})$, $c_{11S}$ is the expression for $c_{11}$ in Reference [17] page 131, the arguments are the radii and the miscentering $x_{ce}$ of the sphere-shell and shell-cavity surfaces, and $C_3$ is the total capacity of the satellite. $c_{11}$ and $-c_{23}$ are also the capacities between the sphere shell and the shell cavity, and when the sphere is exactly centered, $c_{11} = 4\pi\varepsilon_0 ab/(b-a)$ with a similar expression for $-c_{23}$. Thus the number of terms in Smythe necessary for convergence can easily be checked. It is possible to use the theory from a two-sphere system for a three-sphere nested system because of the shielding provided by the shell. Since the charge on the satellite plays no role in the proof-mass forces, the relevant electrostatic energy is $W = \frac{1}{2}[q_1^2/c_{11} - (q_1 + q_2)^2/c_{23}]$ and

$F_1 = \frac{1}{2}(q_1/c_{11})^2 \, dc_{11}/dx_{ce}$ etc.

[5] The charge on the outer shell can be measured by the method used with GP-B [18], i.e. by applying a DC electric field and measuring the electric suspension force necessary to compensate it. Since the sphere is shielded by the shell, the method in [18] cannot be used to measure the charge on the sphere. A charge on the sphere of $q_1$, however, causes a charge on the inner surface of the shell of $-q_1$; and the force between these two charges depends on miscentering of the sphere which acts like a negative spring. Thus the charge on the inner sphere can be measured by forcing an oscillatory motion of the shell with the cavity electrodes and observing the resulting motion of the sphere caused by the coupling of the negative spring. The only known method of proof mass charging from the space environment is from 100 to 200 MeV protons either from solar flares or from trapped protons principally in the South Atlantic Anomaly (SAA). This result is supported by two theoretical papers [16, 18] and by the measurements of the Cactus experiment flown by ONERA in 1975 [19]. The only contrary result is the 1972 DISCOS flight [11] where no proof-mass charging was observed. DISCOS (proof-mass diameter 22 mm, gap 9 mm, and offset 0.7 mm) should have been able to detect charges as small as 200 mV ($5 \times 10^{-13}$ coul or $3 \times 10^6$ electrons). The 750-km polar orbit of DISCOS should have resulted in a charge of 140 mV ($3 \times 10^{-13}$ coul or $2 \times 10^6$ electrons) with each pass through the SAA, so an unknown mechanism was discharging the DISCOS proof mass (perhaps secondary electrons). Its lucky that we didn't know then what we know now, otherwise DISCOS might never have been flown.

Theoretical calculations using the U. S. Navy's Creme96 program (http://crsp3.nrl.navy.mil/creme96) which can calculate geomagnetic shielding, trapped proton flux (using the AP8 models), and proton flux after internal shielding along with the results of Reference [14] show that a 500-km equatorial orbit will experience no charging from 100 to 200 MeV protons. Thus (with the exception of contact potential which can be compensated by the discharge system) it is not known at this time what mechanism would charge the proof masses. Nevertheless the experiment must be capable of measuring proof-mass charges and of discharging them. In the case of a space-station orbit, 380 km and 51.6-degree inclination, the calculations show that a shield of 100 gm/cm² of Invar (about 150 kg in a spherical shell 30 cm in radius) would allow no charging in $10^6$ sec (about 12 days) beyond about 2 mV ($2 \times 10^{-14}$ coul or $10^5$ electrons).

between the sphere and the inner surface of the shell is proportional to $x$ making it behave like a cancellation error (see Table 1). The disturbance from a charge on the shell is carried in the error sources.

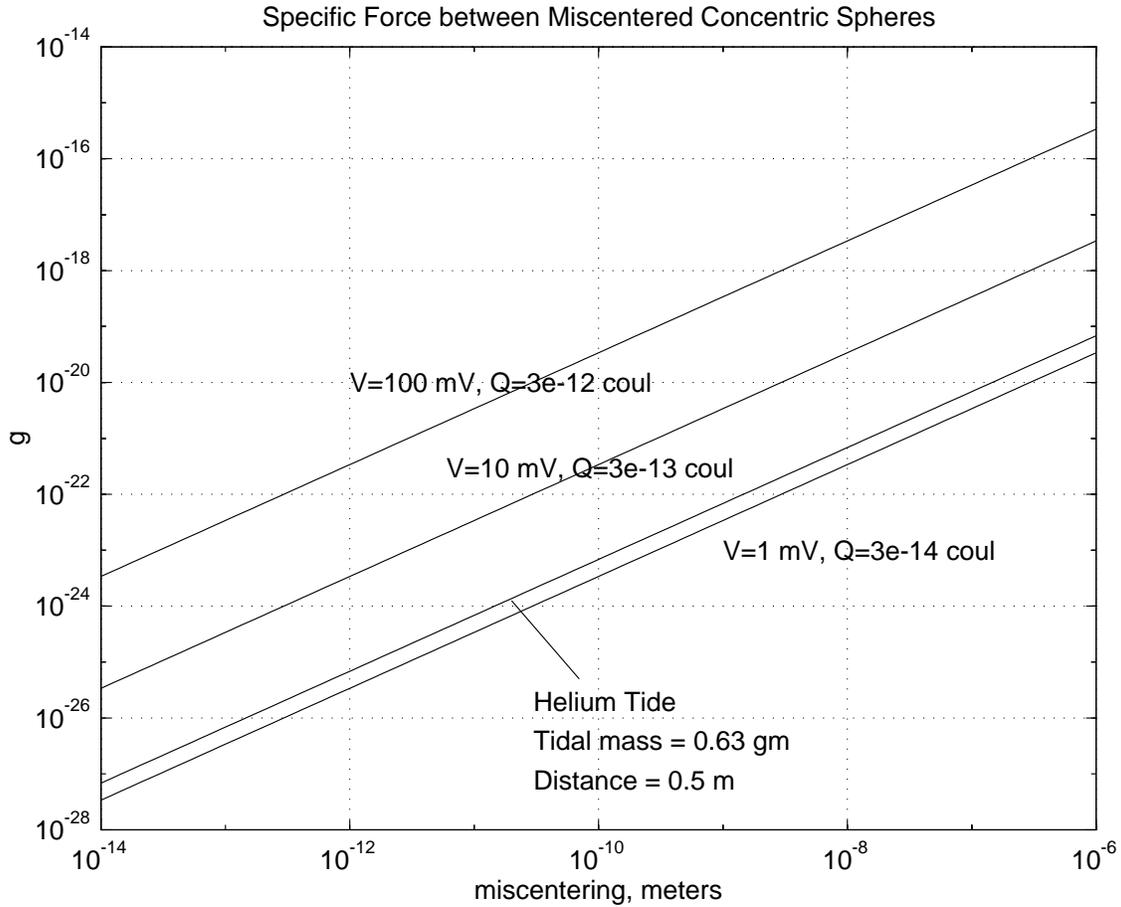

Figure E1. Specific Force on the Proof Masses from Charges and Helium Tides

The applicable value of the miscentering to use in determining the charge or Helium-tide requirements can be found by comparing Figure E1 with Figure 4. It can be seen that although a large value of $f_x$ results in a large miscentering, a large $f_x$ also allows a larger disturbing charge or tide. Thus, for example, a violation of the EP of $10^{-18}$ $g$ would result in a displacement of roughly $10^{-8}$ meters after $10^5$ seconds, but in this case, almost 100 mV of charge potential would be adequate. To the degree that a smaller EP violation results in smaller miscentering, the severity of the charge requirement does not worsen for a more sensitive experiment. It can also be seen from the figure that MiniSTEP-level Helium tides are not a problem for the concentric-sphere experiment.

In the case of the miscentering of the electric shield and forcing electrodes surrounding the proof masses due to imperfect drag-free control, the results of Figure E1 apply to the component at satellite spin frequency. Thus for example for $10^{-20}$ $g$ and 5 mV, the requirement for the drag-free controller is that no error component at spin frequency should exceed about $3 \times 10^{-8}$ meters.

The next largest disturbance at ambient temperature is the radiometer effect which is an unbalance of the cavity pressure caused by a temperature gradient in the proof masses. A hot spot on a proof mass reflects residual gas molecules with greater momentum than the rest of the surface giving a net force. Assuming that one entire side of a proof mass is hotter, this error is given by $(p\pi r^2 / m g_0) A \Delta T / T$. For $p = 10^{-9}$ Torr, $p\pi r^2 / m g_0$ is about $6 \times 10^{-11}$ $g$; and this requires that the roll-frequency component of $\Delta T / T$ times the roll averaging, $A$, be less than $3 \times 10^{-11}$ for an error of $10^{-21}$ $g$. For temperature gradients fixed in the proof masses, proof-mass spin is the source of the roll averaging; and for temperature gradients fixed in the satellite, it comes from satellite spin. With roll averaging of $10^{-6}$, $\Delta T / T$ must be less than $3 \times 10^{-5}$, so that $\Delta T$ is about 0.01 K. This is a relatively easy requirement in a rapidly spinning satellite with one or two cm of cavity insulation [15]. In addition, the large heat capacity of any radiation shield would make this requirement even easier. Differential heating by the UV discharge system and by the transcollimator light beams causes satellite-fixed errors but does not exceed the above requirements.

Misalignment of the z-axis can also cause a radiometer-effect error from its projection into the experiment plane. This effect is roughly fixed in inertial space, but the error is only important if it always acts in the radial direction. Thus there is some roll averaging from the fact that the orbit radius vector rotates once per orbit. This roll averaging, called $A_i$, is limited by the first-half-cycle effect explained in the last paragraph of Section 3.3 so that $A_i$ cannot be smaller than the reciprocal of $\pi$ times the number of revolutions, about $1/160\pi \approx 0.002$ for $10^6$ seconds. An inertial attitude accuracy, $A_{att}$, of $10^{-6}$ radians for the spin axis combined with $A_i$ requires that the temperature difference along $z$ at DC be less than 0.3 K for an error of approximately $10^{-22}$ $g$.

At Helium temperatures the radiometer problem essentially vanishes. This and gas collisions are the principal reasons for doing an experiment a Helium temperatures; although as this paper shows, DC cancellation can give excellent results at room temperature.

Differential radiation pressure at 300 K is also a large effect since it goes as the third power of the temperature times the temperature gradient in the proof masses. For the error calculations, the gradient is assumed to be 0.001 K perpendicular to the spin direction and 0.1 K along the spin axis. $\sigma_{SB}$ is the Stefan-Boltzmann constant.

Differential pressure from out-gassing currents is a similar problem and is more difficult to calculate. Pumping currents arise from the possibility of a vacuum pump removing outgassing products, and they could apply a force to the proof masses. At ambient temperatures for roll-averaging $A = 10^{-6}$ and assuming a differential out-gas pressure of approximately $10^{-5}$ $p$, the disturbance from out-gassing currents would be about $4 \times 10^{-22}$ $g$.

Table E1 shows a list of the disturbances which have not been modeled as random motion. Since there are two proof masses, it was convenient in some cases to have two sets of disturbance calculations, one for each mass. In other cases this was not deemed necessary since we are basically interested in the orders of magnitude of the differential forces. The balance of the errors will be discussed in the order in which they appear in Table E1.

For perfect spheres the gravitational interaction between the sphere and the shell would be zero, and the cancellation from the gravity-gradient masses would be independent of the finite extent of the sphere and shell. All four masses would act as gravitational monopoles, i.e. as points; and the field inside of the shell from the shell would be zero. Because of this, it only necessary to calculate the effects of the error masses arising from the fact that the surfaces are not perfectly spherical and the densities are not perfectly uniform. The error mass was estimated from a volume given by multiplying the polishing error, about $10^{-8}$ meters, by the surface area. Since in the real case, however, the error masses are not all concentrated at a single point but are spread out over the surface of the spheres and are plus and minus with respect to the mean spherical surface; it is assumed that the error masses cancel each other to about one percent, i.e. $A_{dm} = 0.01$.

The dumbbell effect arises because if the mean equatorial figure of either proof mass is not a circle, but an ellipse; the gravitation attraction by the earth will not be given by assuming that the total proof-mass mass is concentrated at its center of mass. If the excess mass, $m_{dum}$, due to the elliptical form is modelled as a dumbbell; there is a second order term in the earth's attraction which is given by $3(r_{pm}^2 / R_e^2) G m_{dum} m_{earth} / R_e^2$. If it is assumed that $m_{dum} \approx m_{err} / 10$, where $m_{err}$ is the total error mass due to surface and density imperfections; this effect gives a disturbance in g's of $3(r_{pm}^2 / R_e^2) m_{err} / 10 m_{1,2}$ which is not reduced by roll averaging since the dumbbell is aligned with the earth twice per proof-mass revolution. Notice that this effect is parallel to the radius vector so that it mimics an EP violation.

The Lorentz force arises from charged proof masses moving at orbital speeds through the earth's magnetic field. In an equatorial orbit, $q \mathbf{v}_O \times \mathbf{B}_e S_h$ causes a disturbance which is in the radial direction, and thus also mimics an EP violation. The only way to reduce this effect is to shield[6] the proof mass and to control the charge. Choosing a polar orbit, however, can reduce this disturbance by one or two orders of magnitude since it would be zero if the earth's dipole were aligned with its spin axis; but since the dipole is tilted approximately 11 deg., the effect is not exactly zero but is reduced by the geometry and the averaging of the earth's rotation.

The subsequent magnetic disturbances are from the force applied to a dipole by a gradient in the magnetic field. Two magnetic fields are considered, one from the earth, $\mathbf{B}_e$, and the other, $\mathbf{B}_i$, a residual constant magnetic field inside of the shield. A dipole moment can arise from two sources, the magnetic susceptibility of the material and eddy currents from the spinning proof masses ($m_{dp} = \pi B_1 r^3 x_{eddy}^2 / 15 \mu_0$). The parameter $x_{eddy}^2 = 2 \mu_0 \sigma \omega r^2$ arises in calculating the magnetic moment from eddy currents in a conducting sphere spinning in a perpendicular magnetic field and is explained in detail in [17 and 4]. The shell, of course, is not a sphere; but it is relatively easy to modify the solution of the boundary value problem for the eddy currents in a sphere described in [17] Chapter 10, p 374 by adding a shell around the sphere. This solution predicts a dipole moment in the spinning shell which is essentially the same as for a sphere, so the spherical solution will be used for both proof masses. For the gradients, $\mathbf{B}_e$ is assumed to vary over a scale equal to the earth's radius and $\mathbf{B}_i$ over the size of the satellite.

---

[6] It is assumed that an AC shielding factor of $10^{-6}$ is possible. The exact degree of shielding is controversial. Two commercial magnetic shield manufacturers, Mumetals and Vacuum Schmeltze, state that they can manufacture multilayer $10^{-6}$ shields; but two GP-B researchers with considerable experience in magnetic shields claim that $10^{-5}$ is about the best that can be done. If $10^{-6}$ cannot be obtained, the deficit can be made up by a tighter control on the charge or by using a polar orbit. The shielding factor at low temperatures is taken from the GP-B results to be $10^{-12}$ [20].

The next three error terms arise from proof-mass and gravity-gradient mass magnetic-dipole interactions. The sphere-shell boundary value solution predicts zero for the sphere-shell magnetic-dipole interaction, but this is unrealistic because the geometry is not perfectly spherical and the magnetic field is not exactly constant. Since the eddy currents, however, are separated in the sphere and the shell; the dipole-dipole interaction can be estimated by calculating the attraction between two dipoles separated by the mean radius of the shell. This estimate is small enough that a more accurate calculation is not needed. The magnetic dipole interaction between the proof and the gravity-gradient masses comes from eddy currents that arise in the gravity-gradient masses because they are rotating in the shielded earth's field. In the perfect case this would cancel at the proof masses, so the error from inexact cancellation is assumed to be one percent, i.e. to come from only one gravity-gradient mass divided by 100. The interaction between the proof masses and the magnetic shield can be calculated by solving the boundary value problem of a slightly offset dipole inside of a sphere of permeability $\mu$. The term $(\mu - 1)/(\mu + 2/3)$ which should appear in the formula is equal to one for large $\mu$ and is omitted.

The z-gravity unbalance term arises because a satellite gravity-gradient error in the z-direction would pull on the proof masses unequally. An attitude error would result in this being projected into the experiment plane. Since this error is roughly fixed in inertial space it is reduced by the assumed attitude error of $10^{-6}$ radians and the partial roll averaging, $A_i$, due to the difference between an inertial vector and the orbit radius vector which is shown above to be about 0.002.

The light pressure errors come from the transcollimator beams and any UV light which might be used to discharge the proof masses.

| Source of Disturbance | Formula for $\delta g / g$ | Auxiliary Formula, Critical Values, Comments, Etc. | Specific Force Error in $g$ | Specific Force Error in $g$ |
|---|---|---|---|---|
| | | | $T = 300$ K $p = 10^{-9}$ Torr | $T = 2$ K $p = 10^{-14}$ Torr |
| | | | | |
| Shell Gravity | $\dfrac{Gm_{err}A}{r_{sh}^2 g_e}$ | $A = 10^{-6}$, $A_{dm} = 0.01$, $m_{err} = 4\pi r_{sh}^2 r_{err} \rho_2 A_{\delta m}$ | 2.0E-23 | 2.0E-23 |
| Dumbbell Effect | $3\dfrac{r_{pm}^2}{R_e^2}\dfrac{m_{err}}{m_{1,2}}$ | $m_{err} = 4\pi r_s^2 r_{err} \rho_2 / 10$ | 4.0E-24 | 4.0E-24 |
| | | | | |
| Lorentz qvBe 1 | $\dfrac{c_{11}V_1 v_O B_e S_h}{m_1 g_e}$ | $V_1 = 5, 1$ mV, $S_h = 10^{-6}, 10^{-12}$ | 1.9E-21 | 3.9E-28 |
| Mag Dipole Bi Chi1 | $\dfrac{\chi_{m1}B_i^2 A}{\mu_0 d_V \rho_1 g_e}$ | $B_i = 10^{-8}, 10^{-10}$ T; $d_V = 1$ meter | 1.6E-25 | 1.6E-29 |
| Mag Dipole Be Chi1 | $\dfrac{\chi_{m1}(B_e S_h)^2}{\mu_0 R_e \rho_1 g_e}$ | $B_e = 2 \times 10^{-5}$ T $A = 10^{-6}$ | 9.9E-32 | 9.9E-44 |
| Mag Dipole Bi Spin 1 | $\dfrac{x_{eddy1}^2 B_i^2 A}{20\mu_0 d_V \rho_1 g_e}$ | $x_{eddy1}^2 = 2\mu_0 \sigma_1 \omega_1 r_1^2 = 6.3 \times 10^{-2}$ | 4.8E-25 | 4.8E-29 |
| Mag Dipole Be Spin 1 | $\dfrac{x_{eddy1}^2 (B_e S_h)^2}{20\mu_0 R_e \rho_1 g_e}$ | $\sigma_1 = 2 \times 10^7$ mho/m | 3.0E-31 | 3.0E-43 |
| | | | | |
| Lorentz qvBe 2 | $\dfrac{c_{22}V_2 v_O B_e S_h}{m_2 g_e}$ | | 8.2E-21 | 1.6E-27 |
| Mag Dipole Bi Chi2 | $\dfrac{\chi_{m2}B_i^2 A}{\mu_0 d_V \rho_2 g_e}$ | | 1.6E-27 | 1.6E-31 |
| Mag Dipole Be Chi2 | $\dfrac{\chi_{m2}(B_e S_h)^2}{\mu_0 R_e \rho_2 g_e}$ | | 1.0E-33 | 1.0E-45 |
| Mag Dipole Bi Spin 2 | $\dfrac{x_{eddy2}^2 B_i^2 A}{20\mu_0 d_V \rho_2 g_e}$ | $x_{eddy2}^2 = 2\mu_0 \sigma_2 \omega_2 r_2^2 = 1.0 \times 10^{-7}$ | 7.4E-31 | 7.4E-35 |
| Mag Dipole Be Spin 2 | $\dfrac{x_{eddy2}^2 (B_e S_h)^2}{20\mu_0 R_e \rho_2 g_e}$ | $\sigma_2 = 10$ mho/m | 4.7E-37 | 4.7E-49 |
| | | | | |
| pm-pm Mag Dipole | $\dfrac{3\mu_0 m_{dp1} m_{dp2}}{4\pi r_{sh}^4 m_{1,2} g_e}$ | $m_{dp1} = 1.6 \times 10^{-10}$ Am$^2$, $m_{dp2} = 1.5 \times 10^{-12}$ Am$^2$ | 1.0E-22 | 1.0E-26 |
| gg-pm Mag Dipole | $\dfrac{6\mu_0 m_{gg-dp} m_{dp1}}{4\pi d_{gg}^4 m_{1,2} g_e 100}$ | $m_{gg-dp} = \dfrac{\pi B_e S_h r_{gg}^3 x_{gg}^2}{15\mu_0}$ | 3.2E-25 | 8.6E-34 |
| pm-Mag Shield Interaction | $\dfrac{6\mu_0 m_{dp1}^2}{4\pi\, r_{shld}^4\, m_{1,2}\, g_e}\left(\dfrac{e_{shld}}{r_{shld}}\right)$ | $e_{shld}$=1 mm, $r_{shld}$=40 cm | 3.6E-28 | 3.6E-32 |
| | | | | |
| Charge 22 | $\dfrac{V_2^2\, dc_{22}\,/\,dx}{2\, m_2\, g_e}$ | $x_{ce}=10^{-7}$ m, $10^{-10}$ m, $V_2$=5, 1 mV | 3.4E-20 | 1.4E-23 |
| | | | | |
| z-Gravity Unbalance | $\dfrac{2G m_{unbal}\, z_{ce}\, A_{att}\, A_i}{d_{unbal}^3\, g_e}$ | $m_{unbal}$=0.1 gm, $d_{unbal}$= 0.1 m, $z_{ce}$=$10^{-7}$ m | 2.7E-28 | 2.7E-28 |
| | | | | |
| zTC Light Pr | $\dfrac{2 P_{beam}\, A_{att}}{c\, m_{1,2}\, g_e}$ | $P_{beam}=10^{-8}$ Watts, $A_{att}=10^{-6}$ rad | 5.1E-26 | 5.1E-26 |
| xyTC Light Pr | $\dfrac{2 P_{beam}\, A}{c\, m_{1,2}\, g_e}$ | $P_{beam}(2K)=10^{-10}$ Watt | 2.5E-23 | 2.5E-23 |
| UV Light Pr | $\dfrac{2 P_{UVbeam}\, A}{c\, m_{1,2}\, g_e}$ | $P_{UVbeam}=6\times10^{-11}$ Watts | 1.6E-25 | 1.6E-25 |
| | | | | |
| Pumping Currents | $\dfrac{p\,\pi\, r_2^2\, A f_{OG}}{m_2\, g_e}$ | Outgassing Pressure Diff. Factor, $f_{OG}=10^{-5}$ | 6.6E-22 | 6.6E-27 |
| Radiometer Effect | $\dfrac{p\,\pi\, r_{1,2}^2\, A\Delta T}{m_{1,2}\, g_e\, T}$ | $p\,\pi\, r_2^2\,/\,m_2\, g_e=$ $4\times10^{-11}$, $10^{-16}$ | 2.2E-22 | 3.3E-25 |
| z-Axis Radiometer Effect | $\dfrac{p\,\pi\, r_{1,2}^2\, A_{att}\, A_i\Delta T_z}{m_{1,2}\, g_e\, T}$ | $A_{att}=10^{-6}$, $A_i$=0.002 $p=10^{-9}$, $10^{-14}$ Torr | 1.3E-23 | 2.0E-26 |
| Differential Radiation Pressure | $\dfrac{\sigma_{SB} T^3 \pi\, r_{1,2}^2\, A 4\Delta T}{c\, m_{1,2}\, g_e}$ | $T=300$, 2 K $\Delta T=0.001$ K | 9.5E-21 | 2.8E-27 |
| Differential z-Axis Radiation Pressure | $\dfrac{\sigma_{SB} T^3 \pi\, r_{1,2}^2\, A_{att}\, A_i 4\Delta T_z}{c\, m_{1,2}\, g_e}$ | $\Delta T_z=0.1$ K | 1.9E-21 | 5.6E-28 |
| | | | | |
| | | RSS of Specific Force Sources, $g$ | 3.7E-20 | 3.5E-23 |

Table E1. Summary of Non-Random Disturbance Accelerations